\documentclass[authoryear,5p]{elsarticle}

\usepackage{lipsum}
\usepackage{amssymb,amsmath}
\usepackage{booktabs}
\usepackage{xspace}
\usepackage[svgnames]{xcolor}
\definecolor{blue}{rgb}{0.3,0.3,0.8}
\definecolor{0206}{rgb}{0.0,0.0,0.0}
\usepackage[colorlinks]{hyperref}

\newcommand{\revision}[2]{{\color{#1}{#2}}}

\newcommand{\degree}{\ifmmode^\circ\else${}^\circ$\fi\relax\xspace}

\newcommand{\dms}[4]{$#1^\mathrm{d}#2^\mathrm{m}#3^\mathrm{s}\!\!.\,#4$}

\journal{Planetary and Space Science}

\begin{document}

\begin{frontmatter}
\title{Relationship between Radar Cross Section and Optical Magnitude based on Radar and Optical Simultaneous Observations of Faint Meteors}

\author[ioa,kiso]{Ryou Ohsawa}
\ead{ohsawa@ioa.s.u-tokyo.ac.jp}

\author[abe]{Akira Hirota}
\author[abe]{Kohei Morita}
\author[abe]{Shinsuke Abe}

\author[irf]{Daniel Kastinen}
\author[irf]{Johan Kero}
\author[irf]{Csilla Szasz}

\author[soken,nms]{Yasunori Fujiwara}

\author[nipr]{Takuji Nakamura}
\author[nipr]{Koji Nishimura}

\author[ioa]{Shigeyuki Sako}
\author[naoj]{Jun-ichi Watanabe}

\author[kiso]{Tsutomu Aoki}
\author[ioa]{Noriaki Arima}
\author[kyoto]{Ko Arimatsu}
\author[ioa,resceu]{Mamoru Doi}
\author[ioa]{Makoto Ichiki}
\author[ism]{Shiro Ikeda}
\author[tohoku]{Yoshifusa Ita}
\author[kyosan,naoj]{Toshihiro Kasuga}
\author[ioa,kiso]{Naoto Kobayashi}
\author[tohoku]{Mitsuru Kokubo}
\author[ioa]{Masahiro Konishi}
\author[okayama]{Hiroyuki Maehara}
\author[ioa]{Takashi Miyata}
\author[kiso]{Yuki Mori}
\author[ism]{Mikio Morii}
\author[ioa]{Tomoki Morokuma}
\author[ioa]{Kentaro Motohara}
\author[ioa]{Yoshikazu Nakada}
\author[jsga]{Shin-ichiro Okumura}
\author[kyosan]{Yuki Sarugaku}
\author[nms]{Mikiya Sato}
\author[resceu]{Toshikazu Shigeyama}
\author[kiso]{Takao Soyano}
\author[ioa,kiso]{Hidenori Takahashi}
\author[tohoku]{Masaomi Tanaka}
\author[kiso]{Ken'ichi Tarusawa}
\author[kipmu,konan]{Nozomu Tominaga}
\author[jsga]{Seitaro Urakawa}
\author[cps]{Fumihiko Usui}
\author[naoj]{Takuya Yamashita}
\author[jaxa]{Makoto Yoshikawa}

\address[ioa]{Institute of Astronomy, Graduate School of Science, The University of Tokyo, 2-21-1 Osawa, Mitaka, Tokyo 181-0015, Japan}
\address[kiso]{Kiso Observatory, Institute of Astronomy, Graduate School of Science, The University of Tokyo 10762-30, Mitake, Kiso-machi, Kiso-gun, Nagano 397-0101, Japan}
\address[abe]{Department of Aerospace Engineering, College of Science \& Technology, Nihon University, 7-24-1 Narashinodai, Funabashi, Chiba 274-8501, Japan}
\address[irf]{Swedish Institute of Space Physics, Box 812, SE-981 28 Kiruna, Sweden}
\address[soken]{SOKENDAI (The Graduate University for Advanced Studies), 10-3 Midoricho, Tachikawa, 190-8518 Tokyo, Japan}
\address[nipr]{National Institute of Polar Research, 10-3, Midori-cho, Tachikawa-shi, Tokyo 190-8518, Japan}
\address[naoj]{National Astronomical Observatory of Japan, 2-21-1 Osawa, Mitaka, Tokyo 181-8588, Japan}

\address[kyoto]{Astronomical Observatory, Graduate School of Science, Kyoto University, Kitashirakawa-oiwake-cho, Sakyo-ku, Kyoto 606-8502, Japan}
\address[resceu]{Research Center for the Early Universe, Graduate School of Science, The University of Tokyo, 7-3-1 Hongo, Bunkyo-ku, Tokyo 113-0033, Japan}
\address[ism]{The Institute of Statistical Mathematics, 10-3 Midori-cho, Tachikawa, Tokyo 190-8562, Japan}
\address[tohoku]{Tohoku University, 6-3 Aramaki, Aoba, Aoba-ku, Sendai, Miyagi 980-8578, Japan}
\address[kyosan]{Department of Physics, Kyoto Sangyo University, Motoyama Kamigamo Kita-ku Kyoto 603-8555 Japan}
\address[okayama]{Okayama Branch Office, Subaru Telescope, National Astronomical Observatory of Japan, NINS, Kamogata, Asakuchi, Okayama, Japan}
\address[nms]{The Nippon Meteor Society}
\address[kipmu]{Kavli Institute for the Physics and Mathematics of the Universe (WPI), The University of Tokyo, 5-1-5 Kashiwanoha, Kashiwa, Chiba 277-8583, Japan}
\address[konan]{Department of Physics, Faculty of Science and Engineering, Konan University, 8-9-1 Okamoto, Kobe, Hyogo 658-8501, Japan}
\address[jsga]{Japan Spaceguard Association, Bisei Spaceguard Center, 1716-3 Okura, Bisei, Ibara, Okayama 714-1411, Japan}
\address[cps]{Center for Planetary Science, Graduate School of Science, Kobe University, 7-1-48 Minatojima- Minamimachi, Chuo-Ku, Kobe, Hyogo 650-0047, Japan}
\address[jaxa]{Japan Aerospace eXploration Agency, 3-1-1 Yoshinodai, Chuo-ku, Sagamihara, Kanagawa 252- 5210, Japan}

\begin{abstract}
Radar and optical simultaneous observations of meteors are important to understand the size distribution of the interplanetary dust. However, faint meteors detected by high power large aperture radar observations, \revision{0206}{which are typically as faint as 10\,mag. in optical,} have not been detected until recently in optical observations, mainly due to insufficient sensitivity of the optical observations. In this paper, two radar and optical simultaneous observations were organized. The first observation was carried out in 2009--2010 using Middle and Upper Atmosphere Radar (MU radar) and an image-intensified CCD camera. The second observation was carried out in 2018 using the MU radar and a mosaic CMOS camera, Tomo-e Gozen, mounted on the 1.05-m Kiso Schmidt Telescope. In total, 331 simultaneous meteors were detected. The relationship between radar cross sections and optical $V$-band magnitudes was well approximated by a linear function. A transformation function from the radar cross section to the $V$-band magnitude was derived for sporadic meteors. The transformation function was applied to about 150,000 meteors detected by the MU radar in 2009--2015, large part of which are sporadic, and a luminosity function was derived in the magnitude range of $-1.5$--$9.5\,\mathrm{mag}$. The luminosity function was well approximated by a single power-law function with the population index  of $r = 3.52{\pm}0.12$. The present observation indicates that the MU radar has capability to detect interplanetary dust of $10^{-5}$--$10^{0}\,\mathrm{g}$ in mass as meteors.

\end{abstract}

\begin{highlights}
\item In total, 331 meteors were detected simultaneously by radar and optically.
\item A correlation between the radar cross section and the optical magnitude is firmly confirmed.
\item The mass range of the meteor detected by MU radar is constrained to about $10^{-5}$--$10^{0}\,\mathrm{g}$.
\end{highlights}

\begin{keyword}
meteors \sep meteoroids \sep interplanetary medium
\end{keyword}

\end{frontmatter}

\section{Introduction}
\label{sec:introduction}
Asteroids and comets are relatively active members in the solar system. They sometimes show cometary activities in the vicinity of the Sun and abruptly eject dust by collision or rapid rotation. The Earth is surrounded by small dust particles generated by such events, which are widely referred to as the interplanetary dust. The mass and size distributions of the interplanetary dust provide important information to understand the origin and evolution of the solar system.

The interplanetary dust grains are steadily colliding with the Earth. The amount of the grains incoming to the Earth is estimated to be about $5$--$300{\times}10^3\,\mathrm{kg}$ per day \citep{plane_cosmic_2012}, which largely consists of grains of $10^{-9}$--$10^{-2}\,\mathrm{g}$ in mass \citep{flynn_extraterrestrial_2002}. The interplanetary dust grains smaller than $10^{-6}\,\mathrm{g}$ have been intensively examined by direct dust detectors onboard spacecrafts \citep{grun_collisional_1985,gruen_ulysses_1992,gurnett_micron-sized_1997,szalay_student_2013}. A detector with a much larger collecting area is, however, required to investigate larger dust grains.

Observing meteors is a method to use the Earth's atmosphere as a huge detector. \revision{0206}{Atoms and molecules in the atmosphere are ionized and excited by an interplanetary dust grain which enters the atmosphere. These processes are interpreted by the theory of thermal ablation \citep{baldwin_ablation_1971,ceplecha_meteor_1998,popova_meteoroid_2004}. Part of the kinetic energy is converted into light. The fraction of the energy converted into light to the lost kinetic energy is called \textit{the luminous efficiency}, which is a key physical quantity to determine the brightness of the meteor. Part of the kinetic energy is also used to ionize surrounding atmospheric molecules. The probability that an atmospheric atom is ionized on a deposition of a single atom from the meteoroid is called \textit{the ionization coefficient}, which determines the electron density of the meteor. There are several attempts to constrain these coefficients observationally \citep[e.g.,][]{verniani_luminous_1965}, experimentally \citep[e.g.,][]{boitnott_light-emission_1972,friichtenicht_determination_1973,thomas_measurements_2016}, and theoretically \citep[e.g.,][]{jones_theoretical_1997}. Radar and optical observations are important to constrain these quantities.}

\revision{0206}{Radar meteors corresponding to extremely faint optical meteors have been detected in the meteor head echo observations with high power and large aperture radar systems such as the European Incoherent Scatter (EISCAT) radars, Advanced Research Projects Agency Long-Range Tracking And Instrumentation Radar (ALTAIR), Middle-and-Upper Atmosphere radar (MU radar), the Middle Atmosphere ALOMAR Radar System (MAARSY), Southern Argentine Agile MEteor Radar (SAAMER), and Poker Flat Incoherent Scatter Radar (PFISR) \citep{pellinen-wannberg_meteor_1994,close_analysis_2000,sato_orbit_2000,kero_first_2011,schult_results_2017,fentzke_latitudinal_2009,sparks_seasonal_2009,janches_interferometric_2014,janches_southern_2015,janches_decade_2019}.} \citet{pellinen-wannberg_meteor_1998} estimated that EISCAT is able to detect meteors as faint as $10\,\mathrm{mag}$ based on the measurements of cross sections and event rates. Thanks to the advances of digital video cameras, such faint meteors have since then been detected optically. Meteors brighter than $4\,\mathrm{mag}$ are routinely collected by large meteor survey networks \citep{jenniskens_meteor_2017}, such as Cameras for All-sky Meteor Surveillance \citep[CAMS;][]{jenniskens_cams:_2011}, European viDeo MeteOr Network Database \citep[EDMOND;][]{kornos_database_2013}, and SonotaCo network \citep{kanamori_meteor_2009}. Recently, high-sensitive optical camera systems make it possible to observe meteors as faint as ${\sim}10\,\mathrm{mag}$ in the optical regime \citep{nishimura_high_2001,weryk_canadian_2013,ohsawa_luminosity_2019}.

Since every observation method is subject to biases and errors, simultaneous observations of the same meteors using different methods are important. Part of previous simultaneous observations are described here. \citet{fujiwara_simultaneous_1995} carried out simultaneous observations with the MU radar and three video cameras with image intensifiers. In two observation runs, they detected 19 simultaneous meteors and suggested a log-linear relationship between the radar received power and the optical magnitude for Geminids. \citet{nishimura_high_2001} detected 35 simultaneous meteors during two nights using the MU radar and an image-intensified CCD video camera. They showed that the radar received power changed partly in synchronization with the optical brightness. They also confirmed a log-linear relationship between the radar received power and the optical magnitude for sporadic meteors. \revision{0206}{\citet{michell_simultaneous_2010} detected seven meteors in optical out of 338 meteors observed with PFISR, and confirmed that a similar positive correlation between the optical brightness and the back-scattered radar power. Using a combination of SAAMER and an EMCCD video camera, \citet{michell_simultaneous_2015} obtained 6 meteors simultaneously by radar and optically and showed that the meteoroid masses independently estimated from the radar and optical observations were roughly consistent with each other. \citet{michell_simultaneous_2019} carried out optical and dual-frequency radar observations with the Arecibo radar and an EMCCD camera. In total, 19 meteor events were detected simultaneously in the three methods. No apparent correlation between the optical mass and the radar signal-to-noise ratios was, however, confirmed. Since not a small fraction of the meteors were detected in side lobes, they presumed that the radar cross sections could be underestimated, resulting in a possible artificial bias.} \citet{campbell-brown_photometric_2012} carried out observations with EISCAT and two image-intensified CCD video cameras and detected 4 meteors whose orbits were determined both by radar and optically. They confirmed that the photometric and ionization masses were consistent within estimated errors. \revision{0206}{\citet{weryk_simultaneous_2013,weryk_simultaneous_2012} used the Canadian Meteor Orbit Radar (CMOR) and multiple image-intensified CCD video cameras and detected about ${\sim}500$ simultaneous meteors in the magnitude range of about $0$--$5\,\mathrm{mag}$. They derived the relationship between the electron line density and the photon radiant power. The dependency of the ionization coefficient to the luminous efficiency ratio on the speed of meteors was observationally constrained. The luminous efficiency they derived indicated that the meteoroid mass flux in the range $10^{-5}$--$10^{-2}\,\mathrm{g}$ could be lower than previously estimated.} \citet{brown_simultaneous_2017} carried out observations with MAARSY and two image-intensified video cameras. They detected more than 100 simultaneous meteors whose orbits were constrained both by radar and optically. The magnitudes of the detected meteors ranged from $0$--$7\,\mathrm{mag}$. They found a clear trend where brighter meteors showed higher peak radar cross sections. More complete overviews are found in \citet{weryk_simultaneous_2012,weryk_simultaneous_2013} and \citet{brown_simultaneous_2017}.

Although there have been a number of radar and optical simultaneous observations in literature, meteors fainter than ${\sim}6\,\mathrm{mag}$ have not been fully investigated. This is partly because the performances of optical camera systems are limited. A combination of a large aperture, high-speed and continuous imaging, and a large field-of-view is required to detect faint meteors optically. Recently, \citet{ohsawa_luminosity_2019} carried out observations of meteors as faint as ${\sim}10\,\mathrm{mag}$ with a mosaic CMOS camera, Tomo-e PM, mounted on the 1.05-m Kiso Schmidt telescope. A combination of a large and wide-field telescope and a mosaic CMOS camera is preferable to detect faint meteors. Here, we report two simultaneous observations of faint sporadic meteors. \revision{0206}{The observations are intended to confirm the trends between the radar cross section in the meteor head echo observation and the optical magnitude suggested in previous studies \citep{nishimura_optical_2002,michell_simultaneous_2015,brown_simultaneous_2017} with a large number of samples and a wide magnitude range.} In the first observation run, observations were carried out with the MU radar and an image-intensified CCD video camera in 2009--2010. The second observation run were carried out with the MU radar and Tomo-e Gozen, which is a successor of Tomo-e PM. The paper is organized as follows; Details of the observations are described in Section~\ref{sec:observations}. Statistical properties of detected meteors are presented in Section~\ref{sec:results}. In Section~\ref{sec:discussion}, the relationship between the radar cross section and the optical brightness and the luminosity function of sporadic meteors are discussed, and then Section~\ref{sec:conclusion} summarizes this paper.

\section{Observations}
\label{sec:observations}
\newcommand{\runA}{ICCD09\xspace}
\newcommand{\runB}{KISO18\xspace}

\subsection{Observations with the MU radar and Intensified CCD camera}

\begin{table*}
\centering
\caption[Observation Summary in 2009--2010]{Summary of Observations with the MU radar and ICCD camera in 2009--2010}
\label{tab:obs:summary:iccd}
\small
\begin{tabular}{ll}
\toprule
Date & Sep. 24--26, Oct. 19--21, Nov. 8, and Dec. 13--14 in 2009 \\
& Mar. 11, Aug. 12, Sep. 13, and Dec. 14 in 2010 \\
Radar System & Middle and Upper Atmosphere Radar (46.5 MHz) \\
Optical System & Canon 200\,mm F/1.8 and Hamamatsu CCD camera (ORCA-05G) \\
Video Frame Rate & 29.97\,Hz \\
Field of View & ${\sim}6\degree$ in diameter (radar),
$9.9\degree{\times}10.9\degree$ (optical) \\
\bottomrule
\multicolumn{2}{p{.8\linewidth}}{}
\end{tabular}

\centering
\caption[Observation Summary in 2018]{Summary of Observations with the MU radar and Tomo-e Gozen in 2018}
\label{tab:obs:summary:tomoe}
\small
\begin{tabular}{ll}
\toprule
Date & Apr. 18--22 in 2018 (11:00--20:00 UT, 36 hours in total) \\
Radar System & Middle and Upper Atmosphere Radar (46.5 MHz) \\
Optical System & 1.05-m Kiso Schmidt telescope and Tomo-e Gozen (Q1) \\
Video Frame Rate & 2.0\,Hz \\
Field of View & ${\sim}6\degree$ in diameter (radar),
$20{\times}~ 39.7'{\times}22.4'$ (optical) \\
\bottomrule
\multicolumn{2}{p{.8\linewidth}}{}
\end{tabular}
\end{table*}

The first observation run was carried out in 2009 and 2010 (hereafter, referred as to \runA). Specifications of the observations are summarized in Table~\ref{tab:obs:summary:iccd}. We had several observations in 2009 and 2010. Radar observations were carried out with Middle and Upper Atmosphere Radar (hereafter, MU radar) in the Shigaraki MU Observatory\footnote{Shigaraki MU Observatory is located at \dms{+34}{51}{14}{5}\,N and \dms{+136}{06}{20}{24}\,E (WGS84).} of Research Institute for Sustainable Humanosphere (RISH), Kyoto University. The MU radar was operated in the general head echo mode \citep{kero_first_2011}. Optical observations were carried out with an image-intensified CCD camera made by Hamamatsu equipped with a Canon 200\,mm F/1.8 lens. The optical camera system was set up in Shigaraki. The camera was pointed toward zenith and continuously monitored the sky at 29.97\,Hz. GPS time stamps were imprinted in the video data.

The radar data were reduced using a standard data reduction process of the MU radar \citep{kero_first_2011,kero_2009-2010_2012,kero_meteor_2012}. The three dimensional trajectory and the radar cross section (RCS) \revision{0206}{along the trajectory} of each meteor were obtained. Meteors in the optical data were detected with a time shifted motion capture software, UFOCapture\footnote{UFOCaptrueV2 (ver 2.24) in \url{http://sonotaco.com/soft/e_index.html}}. The optical trajectories and the magnitudes of the meteors were derived with a post processing tool, UFOAnalyzer\footnote{UFOAnalyzerV2 (ver 2.44) in \url{http://sonotaco.com/soft/e_index.html}}.

\subsection{Observations with the MU radar and Tomo-e Gozen}

The second observations were carried out in April, 2018 (hereafter, referred as to \runB). Radar observations were also carried out with the MU radar in the same setting as in \runA. Optical observations were carried out with a mosaic CMOS camera, Tomo-e Gozen, mounted on the 1.05-m Schmidt Telescope in Kiso Observatory\footnote{Kiso Observatory is located at \dms{+35}{47}{38}{7}\,N and \dms{+137}{37}{42}{2}\,E (WGS84).} of the Institute of Astronomy, the University of Tokyo. Specifications of the observations are summarized in Table~\ref{tab:obs:summary:tomoe}. Tomo-e Gozen is equipped with 84 CMOS image sensors of $2000{\times}1128$ pixels in size. The field-of-view is, in total, as large as about $20\,\mathrm{sq.\,degree}$, and Tomo-e Gozen is able to monitor the sky up to at 2\,Hz \citep{sako_development_2016,sako_tomo-e_2018,kojima_evaluation_2018}. The readout of the image sensor is synchronized with the GPS time and the time stamp of Tomo-e Gozen is as accurate as 0.2\,ms \citep{sako_tomo-e_2018}. Observations are carried out in a clear filter and a nominal limiting magnitude for stars is about 18\,mag, which is corresponding to about 12\,mag for meteors \citep{ohsawa_luminosity_2019}. At the time of the observations, only one quadrant of the camera was available and one sensor was not operating. Thus, the observations were carried out with 20 CMOS sensors (${\sim}4.8\,\mathrm{sq.\,degree}$ in total). The Kiso Schmidt telescope was pointed toward the sky 100\,km above the MU radar. Since the telescope tracked the sky, the direction of the telescope was adjusted every 3 minutes. Thus, the length of each video is 3 minutes. Kiso Observatory is located about 173\,km distant from the Shigaraki MU Observatory. The elevation angle of the telescope was about 30\degree and the distance between the telescope and the volume monitored by the MU radar was about 200\,km.

\begin{figure*}
\centering
\includegraphics[width=\linewidth]{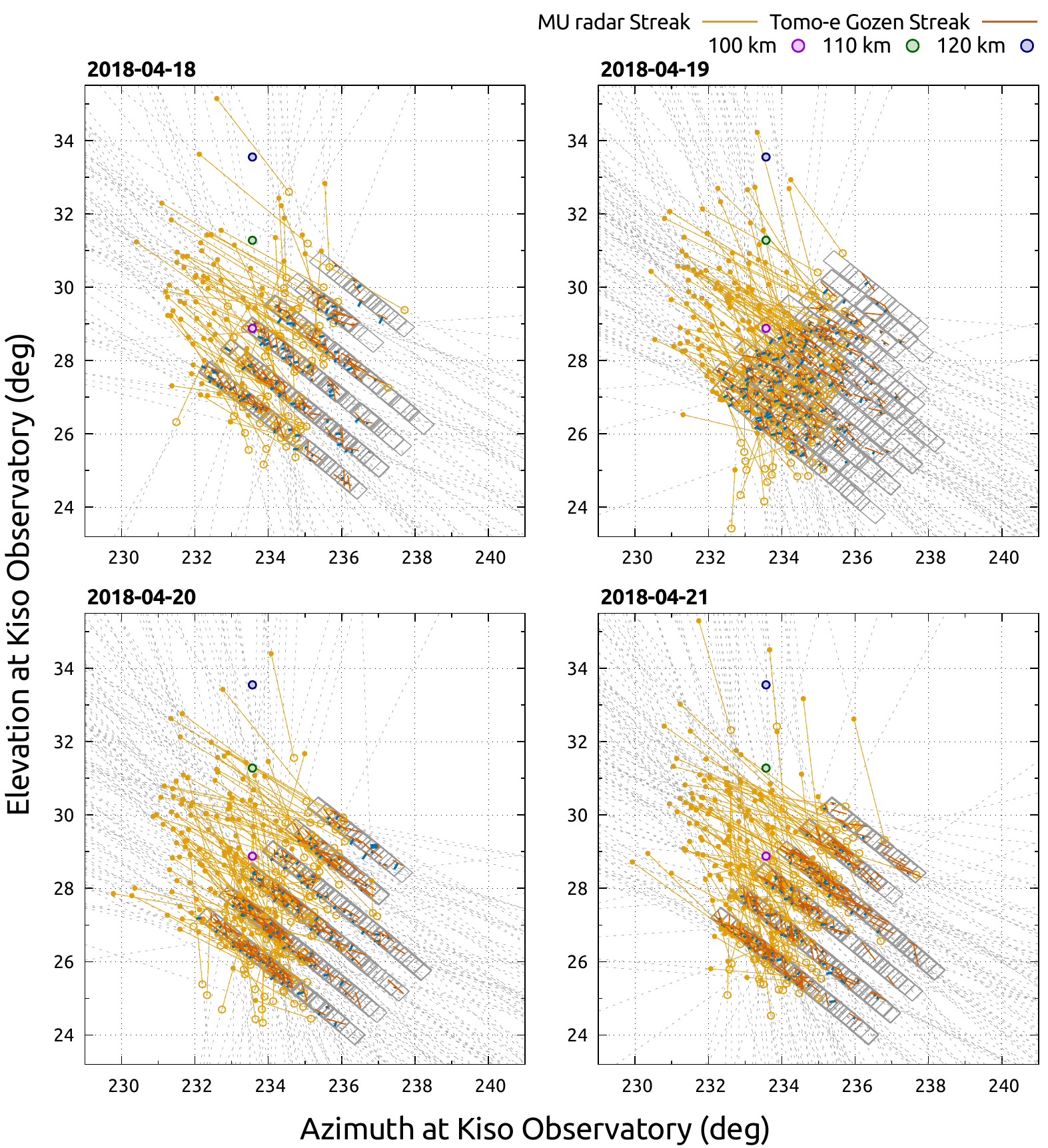}
\caption[Meteors at Kiso Observatory]{The meteors detected in \runB are projected onto the sky from Kiso Observatory in the elevation and azimuth coordinates. Each panel illustrates the meteors detected in a night. The orange segments indicate the meteors detected by the MU radar; The filled and empty circles are, respectively, the first and last detection points. The gray dashed lines are the extensions of the MU trajectories for reference. The gray rectangles are the fields-of-view of Tomo-e Gozen. The red segments indicate the meteors detected by Tomo-e Gozen. The blue segments indicate the distances between the meteor segments detected by the MU radar and Tomo-e Gozen. The violet, green, and navy circles respectively indicate the center of the field-of-view of the MU radar at $100$, $110$, and $120\,\mathrm{km}$ in altitude.}
\label{fig:results:trajectory}
\end{figure*}

The radar data were reduced in the same manner as in \runA. \revision{0206}{The three dimensional trajectory and the RCS of each meteor were obtained.} The optical data were reduced in a standard astronomical data reduction procedure for imaging observations: a dark frame was subtracted and a flat frame correction was applied. Meteors in the optical video data were extracted with an algorithm based on the Hough transformation \citep{ohsawa_development_2016,ohsawa_luminosity_2019}. The detected events are summarized in Figure~\ref{fig:results:trajectory}. The meteors detected by the MU radar are shown by the orange segments, while those detected by Tomo-e Gozen are shown by the red segments. The gray rectangles indicate the fields-of-view of Tomo-e Gozen. A number of rectangles appear in Figure~\ref{fig:results:trajectory} since the telescope was moved frequently. The violet, green, and navy circles in Figure~\ref{fig:results:trajectory} indicate the center of the field-of-view of the MU radar at different altitudes for reference. This indicates that the fields-of-view of Tomo-e Gozen were, however, slightly displaced due to a miscalculation.

\subsection{Archival Observations from the MU radar Meteor Head Echo database}
Meteors detected by the MU radar were retrieved from an archival data, in order to discuss the luminosity function of faint meteors. Part of data were already published in \citep{kero_2009-2010_2012,kero_first_2011} and available in the MU Radar Head Echo Database (MURMHED). All the data were reduced along with \citet{kero_meteor_2012}. The observations were carried out during 2009--2015 and the total observing time was 845.8\,hour. The archive contains 157043 meteor events in total. \revision{0206}{The contributions from meteor showers were removed based on the D-criterion\footnote{\revision{0206}{The calculation was based on the orbital elements of the 112 established meteor showers issued in the IAU Meteor Data Center on February 17, 2020.}}, which is a criterion to determine whether a meteor belongs to a stream or not \citep{southworth_statistics_1963}. The threshold of the D-criterion was set to 0.2. This threshold is relatively weak compared to previous studies which identify meteor showers by the D-criterion \citep[e.g.,][]{jenniskens_cams_2016}, but reasonable to roughly estimate the contributions from meteor showers \citep{andreic_results_2014,segon_results_2014,gural_results_2014,andreic_ten_2013}. Some sporadic meteors may be wrongly removed due to the weak threshold, but the result will not be affected. The removed meteors were mainly composed of Geminids (1034), Orionids (2049), and $\eta$ Aquariids (1900). Finally, 150080 meteors were extracted as sporadic meteors.} The archive was complied with several campaign observations. The number of events per month is listed in Table~\ref{tab:observation:murmhed}. The table indicate that the observations were biased toward meteors observed in October and December.

\begin{table*}
\centering
\caption{The Number of Sporadic Meteor Events in MURMHED}
\label{tab:observation:murmhed}
\begin{tabular}{cllllllllllll}
\toprule
\multicolumn{1}{@{\quad}c@{\quad}}{Year}
& \multicolumn{1}{c}{Jan.}
& \multicolumn{1}{c}{Feb.}
& \multicolumn{1}{c}{Mar.}
& \multicolumn{1}{c}{Apr.}
& \multicolumn{1}{c}{May}
& \multicolumn{1}{c}{Jun.}
& \multicolumn{1}{c}{Jul.}
& \multicolumn{1}{c}{Aug.}
& \multicolumn{1}{c}{Sep.}
& \multicolumn{1}{c}{Oct.}
& \multicolumn{1}{c}{Nov.}
& \multicolumn{1}{c}{Dec.} \\\midrule
2009
& \multicolumn{1}{c}{---}
& \multicolumn{1}{c}{---}
& \multicolumn{1}{c}{---}
& \multicolumn{1}{c}{---}
& \multicolumn{1}{c}{---}
& \multicolumn{1}{c}{3819}
& \multicolumn{1}{c}{5465}
& \multicolumn{1}{c}{---}
& \multicolumn{1}{c}{5419}
& \multicolumn{1}{c}{9071}
& \multicolumn{1}{c}{4925}
& \multicolumn{1}{c}{4632} \\
2010
& \multicolumn{1}{c}{3398}
& \multicolumn{1}{c}{2788}
& \multicolumn{1}{c}{2000}
& \multicolumn{1}{c}{2205}
& \multicolumn{1}{c}{2479}
& \multicolumn{1}{c}{3064}
& \multicolumn{1}{c}{3570}
& \multicolumn{1}{c}{8520}
& \multicolumn{1}{c}{4441}
& \multicolumn{1}{c}{21467}
& \multicolumn{1}{c}{4856}
& \multicolumn{1}{c}{8314} \\
2011
& \multicolumn{1}{c}{---}
& \multicolumn{1}{c}{---}
& \multicolumn{1}{c}{---}
& \multicolumn{1}{c}{---}
& \multicolumn{1}{c}{---}
& \multicolumn{1}{c}{---}
& \multicolumn{1}{c}{---}
& \multicolumn{1}{c}{---}
& \multicolumn{1}{c}{---}
& \multicolumn{1}{c}{6087}
& \multicolumn{1}{c}{---}
& \multicolumn{1}{c}{---} \\
2012
& \multicolumn{1}{c}{---}
& \multicolumn{1}{c}{---}
& \multicolumn{1}{c}{---}
& \multicolumn{1}{c}{---}
& \multicolumn{1}{c}{---}
& \multicolumn{1}{c}{---}
& \multicolumn{1}{c}{---}
& \multicolumn{1}{c}{---}
& \multicolumn{1}{c}{---}
& \multicolumn{1}{c}{10930}
& \multicolumn{1}{c}{---}
& \multicolumn{1}{c}{---} \\
2013
& \multicolumn{1}{c}{3298}
& \multicolumn{1}{c}{2231}
& \multicolumn{1}{c}{1708}
& \multicolumn{1}{c}{---}
& \multicolumn{1}{c}{---}
& \multicolumn{1}{c}{---}
& \multicolumn{1}{c}{---}
& \multicolumn{1}{c}{---}
& \multicolumn{1}{c}{---}
& \multicolumn{1}{c}{---}
& \multicolumn{1}{c}{---}
& \multicolumn{1}{c}{13332} \\
2014
& \multicolumn{1}{c}{3300}
& \multicolumn{1}{c}{---}
& \multicolumn{1}{c}{---}
& \multicolumn{1}{c}{---}
& \multicolumn{1}{c}{136}
& \multicolumn{1}{c}{---}
& \multicolumn{1}{c}{---}
& \multicolumn{1}{c}{---}
& \multicolumn{1}{c}{---}
& \multicolumn{1}{c}{---}
& \multicolumn{1}{c}{---}
& \multicolumn{1}{c}{6477} \\
2015
& \multicolumn{1}{c}{---}
& \multicolumn{1}{c}{---}
& \multicolumn{1}{c}{---}
& \multicolumn{1}{c}{2148}
& \multicolumn{1}{c}{---}
& \multicolumn{1}{c}{---}
& \multicolumn{1}{c}{---}
& \multicolumn{1}{c}{---}
& \multicolumn{1}{c}{---}
& \multicolumn{1}{c}{---}
& \multicolumn{1}{c}{---}
& \multicolumn{1}{c}{---} \\
\midrule
total
& \multicolumn{1}{c}{9996}
& \multicolumn{1}{c}{5019}
& \multicolumn{1}{c}{3708}
& \multicolumn{1}{c}{4353}
& \multicolumn{1}{c}{2615}
& \multicolumn{1}{c}{6883}
& \multicolumn{1}{c}{9035}
& \multicolumn{1}{c}{8520}
& \multicolumn{1}{c}{9860}
& \multicolumn{1}{c}{47555}
& \multicolumn{1}{c}{9781}
& \multicolumn{1}{c}{32755} \\
\bottomrule
\end{tabular}
\end{table*}

\section{Results}
\label{sec:results}
\subsection{Simultaneous Meteors of \runA}
We define ``a simultaneous meteor'' as a meteor observed both by radar and optically. In case of \runA, simultaneous meteors were identified based on the trajectories and timings of meteors. When meteors were detected both by radar and optically and the time separation was within 0.5\,s, the meteors were considered as identical. Finally 145 meteors were identified as the simultaneous events.

Some observations were scheduled when meteor showers were active. Meteors belonging to showers were removed based on the D-criterion. The threshold for the D-criterion was set to 0.2 similarly as descried above. The present dataset \revision{0206}{possibly contained 5 Southern Taurids, 13 Geminids, 17 Orionids, 1 Andromedids, 2 December Monocerotids, 1 Comae Berenicids, 2 $\varepsilon$ Geminids, 16 $\eta$ Aquariids, 1 October Capricornids, and 1 November Orionids. The remaining 103 meteors were considered to be sporadic. In the following sections, the 103 sporadic meteors were investigated.}

The brightness of the \runA meteors were calibrated against the $V$-band magnitudes in SKY2000 catalog version 4 \citep{myers_vizier_2001} using UFOAnalyzer. \revision{0206}{The correction of the color term was not applied. Note that the derived magnitudes could be affected by an amount depending on the unknown spectral character of the meteors.} The light curve of each meteor was derived. \revision{0206}{About 20\% of the meteors showed light bursts with amplitudes of larger than $1.5,\mathrm{mag}$ in their light curves, possibly attributed to fragmentation. Such sudden bursts are not necessarily accompanied by variations in the RCS \citep[e.g.,][]{brown_simultaneous_2017}. The meteors obtained in \runB were rarely affected by such bursts since they were observed in a narrow fields-of-view. To enable a fair comparison with the magnitude in \runB, the magnitude averaged over the streak was adopted as the representative value of each meteor, instead of the peak magnitude, since the latter could be affected by fragmentation. Since the bursts were sufficiently short, the variability in the brightness had little effect on the average brightness values.} Finally, the observed magnitudes were converted into the meteor absolute magnitudes.

\subsection{Simultaneous Meteors of \runB}
To extract simultaneous meteors, we first sifted the optical meteors detected within ${\pm}1$ frames (1.5\,s) of the time stamp of each meteor detected by radar. Then, the meteors detected by radar were projected onto the sky from Kiso Observatory (Figure~\ref{fig:results:trajectory}). The candidates of the simultaneous meteors were selected based on the differences in direction and the separations; The angle between the radar (orange) and optical (red) trajectories should be smaller than 2.5\degree and the separation angle of the two trajectories (the blue segment in Figure~\ref{fig:results:trajectory}) should be smaller than 0.25\degree. When the same meteor was detected in multiple detectors, the brightest segment was adopted as the representative one. The total number of unique simultaneous meteors was 485 in the four nights. As shown in Figure~\ref{fig:results:trajectory}, the pointing of the telescope was displaced. The radar and optical observations sometimes traced completely different portions of the trajectory. The magnitudes of such meteors could be erroneous. Thus, we removed about 250 simultaneous meteors whose optical trajectories were not completely covered by the radar observations. \revision{0206}{A possible contribution from meteor showers were removed based on the D-criterion. In total, 5 meteors were removed.} Finally, the sample size was reduced to 228, \revision{0206}{which accounted for about 8\% of the total number of the meteors whose radar trajectories crossed the fields-of-view of Tomo-e Gozen}. In the following sections, the 228 simultaneous sporadic meteors were investigated.

The meteors in \runB were observed at 2\,Hz. Thus, the meteors were captured as streaks. \revision{0206}{We calculated the magnitudes of the meteors, following the method used in \citet{iye_suprimecam_2007}, to derive the brightness of the meteors from the detected streaks. The brightness of the meteor was estimated by $I_v\,v\,T$, where $I_v$ is the line intensity averaged along the streak, $v$ is the angular velocity of the meteor, and $T$ is the exposure time of each video frame.} In \citet{iye_suprimecam_2007}, the meteor speeds $v$ were uniformly assumed to be $10\degree\,\mathrm{s^{-1}}$. The magnitudes of meteors were also derived based on the same assumption in \citet{ohsawa_luminosity_2019}. This assumption was a major source of the uncertainty in those studies. In the \runB observation, the projected motion of each meteor was directly derived from the MU radar observation. The magnitudes in the present research were little affected by the uncertainty in the meteor speed. The magnitudes of the meteors were first calibrated against the $V$-band magnitudes of the UCAC4 catalog \citep{zacharias_fourth_2013}. \revision{0206}{The color term correction was not applied. Note that the derived magnitudes could be affected by an amount dependeng on the unknown spectral character of the meteors.} Then, the magnitudes were finally converted into the meteor absolute magnitudes using the distance between the telescope and the meteors.

\subsection{Statistics of Simultaneous Meteors}

\begin{figure*}
\centering
\includegraphics[width=0.9\linewidth]{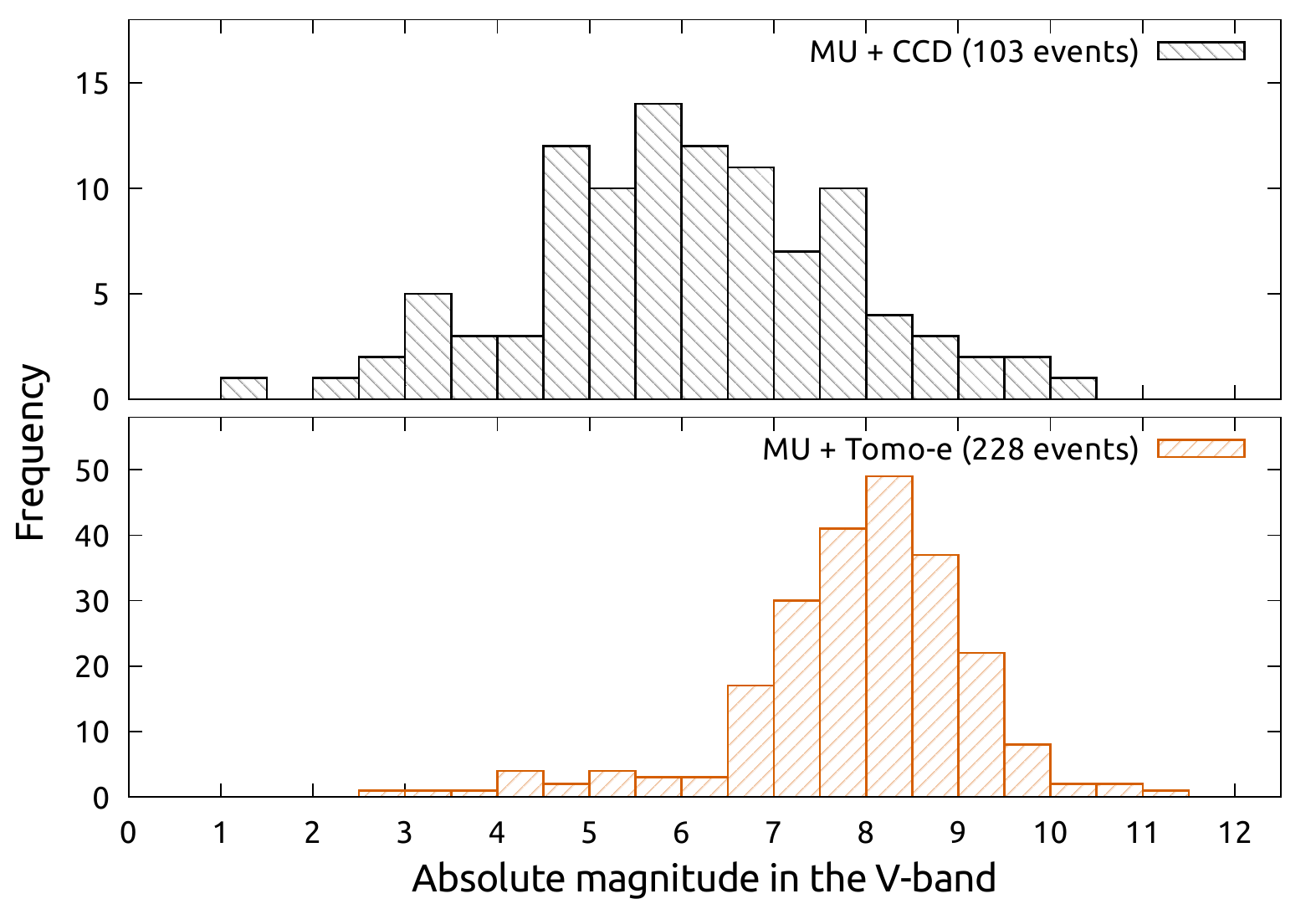}
\caption[Magnitude distributions of the simultaneous meteors]{Magnitude distributions of the meteors simultaneously detected by radar and optically. The top panel shows the magnitude distribution of the observations in 2009--2010 (\runA), while the bottom panel illustrates that of the observations in 2018 (\runB).}
\label{fig:results:histogram}
\end{figure*}

The distributions of the absolute magnitude is shown in Figure~\ref{fig:results:histogram}. The top panel shows the distribution of \runA, while the bottom panel does that of \runB. \revision{0206}{The brightest meteor in \runA was $1.4\,\mathrm{mag}$, while the faintest meteor was $10.0\,\mathrm{mag}$. The magnitudes at 25-, 50-, and 75-percentiles were $5.0$, $6.1$, and $7.2\,\mathrm{mag}$, respectively. The brightest meteor in \runB was about $2.8\,\mathrm{mag}$, while the faintest meteor was $11.1\,\mathrm{mag}$. The magnitudes at 25-, 50-, and 75-percentiles were $7.4$, $8.1$, and $8.6\,\mathrm{mag}$, respectively.} The detected meteors in \runB were typically about $1.8\,\mathrm{mag}$ fainter than those in \runA, in spite of the larger distance between the radar and optical observation sites. This is simply attributed to the high sensitivity of Tomo-e Gozen.

\begin{figure*}
\centering
\includegraphics[width=0.9\linewidth]{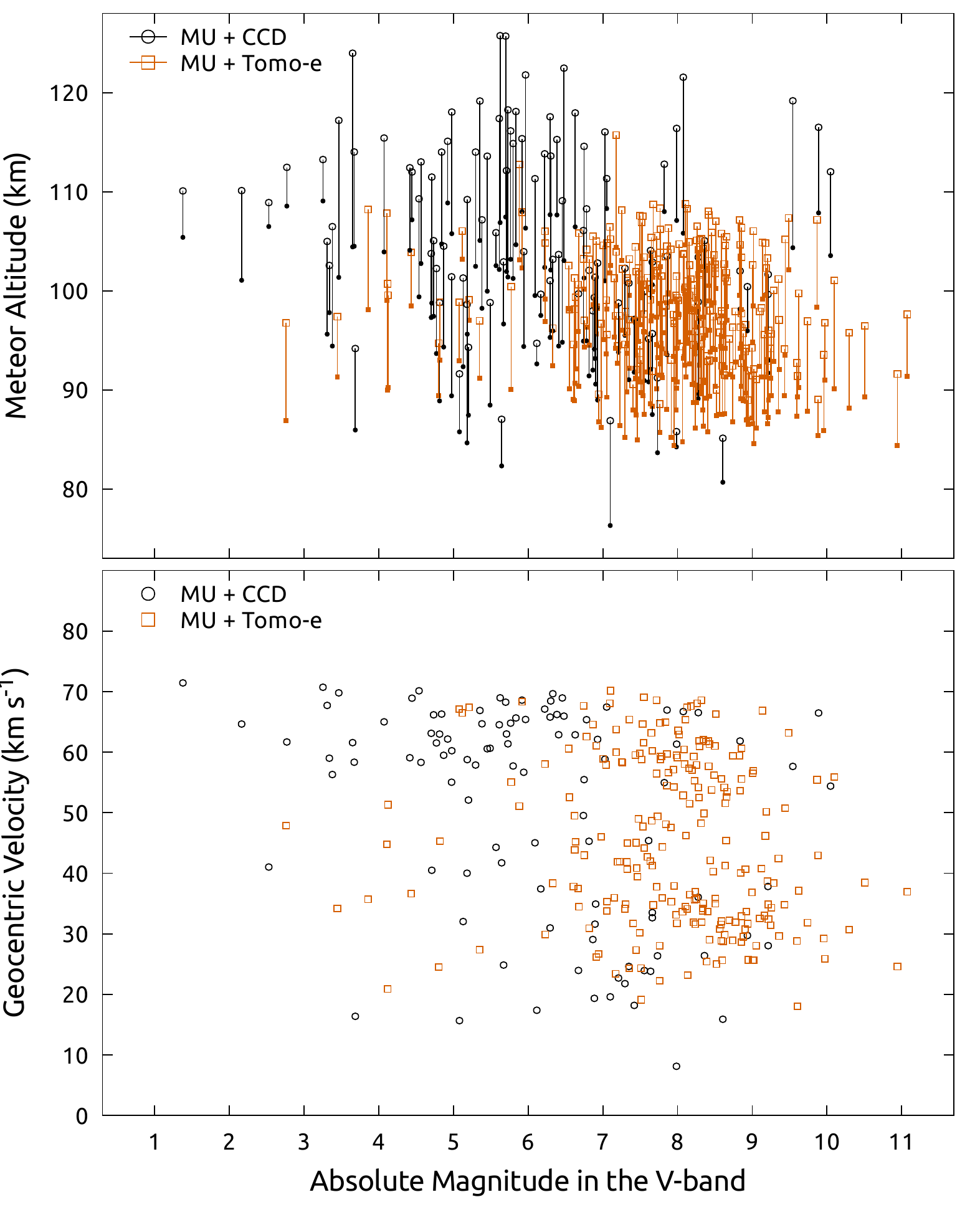}
\caption[Height and Velocity of the simultaneous meteors]{The altitude and the geocentric velocity are shown against the optical magnitude in the $V$-band. The data of \runA are shown by the black circle symbols, while those of \runB are shown by the red square symbols. The top panel illustrates the observed heights in km. The empty and filled symbols are respectively the highest and lowest altitudes measured by the MU radar. The geocentric velocity in $\mathrm{km\,s^{-1}}$ is shown in the bottom panel.}
\label{fig:results:stats}
\end{figure*}

The top panel of Figure~\ref{fig:results:stats} displays the altitudes of the meteors against the optical absolute magnitudes. \revision{0206}{No apparent dependency of the altitude on the optical magnitude was recognized. But this does not mean the distribution of the meteoroid mass was uniform along the altitude. It should be noted that a limiting meteoroid mass is smaller at a higher altitude for a magnitude-limited sample \citep[e.g., \textit{see} Figure 4 in][]{ceplecha_meteor_1998}.} The altitudes of the \runB meteors are generally lower than those of the \runA meteors. As shown in Figure~\ref{fig:results:trajectory}, the pointing of the telescope in \runB was displaced downward in elevation. The fields-of-view of Tomo-e Gozen were set below about 30\degree in elevation, corresponding to about $100\,\mathrm{km}$ in altitude above the MU radar. Since the differences between the highest and lowest altitudes is typically about $10\,\mathrm{km}$ in Figure~\ref{fig:results:stats}, the meteors below $110\,\mathrm{km}$ were selectively detected as simultaneous meteors in \runB. Thus, the difference in the altitude distributions is explained by the observation bias. Consequently, the present samples contained few meteors which were fainter than about $7\,\mathrm{mag}$ and whose altitudes were higher than $100\,\mathrm{km}$. The geocentric velocities are plotted against the optical magnitudes in the bottom panel of Figure~\ref{fig:results:stats}. No apparent trend is confirmed. The distribution of the geocentric velocity is bimodal.

\begin{figure*}
\centering
\includegraphics[width=0.9\linewidth]{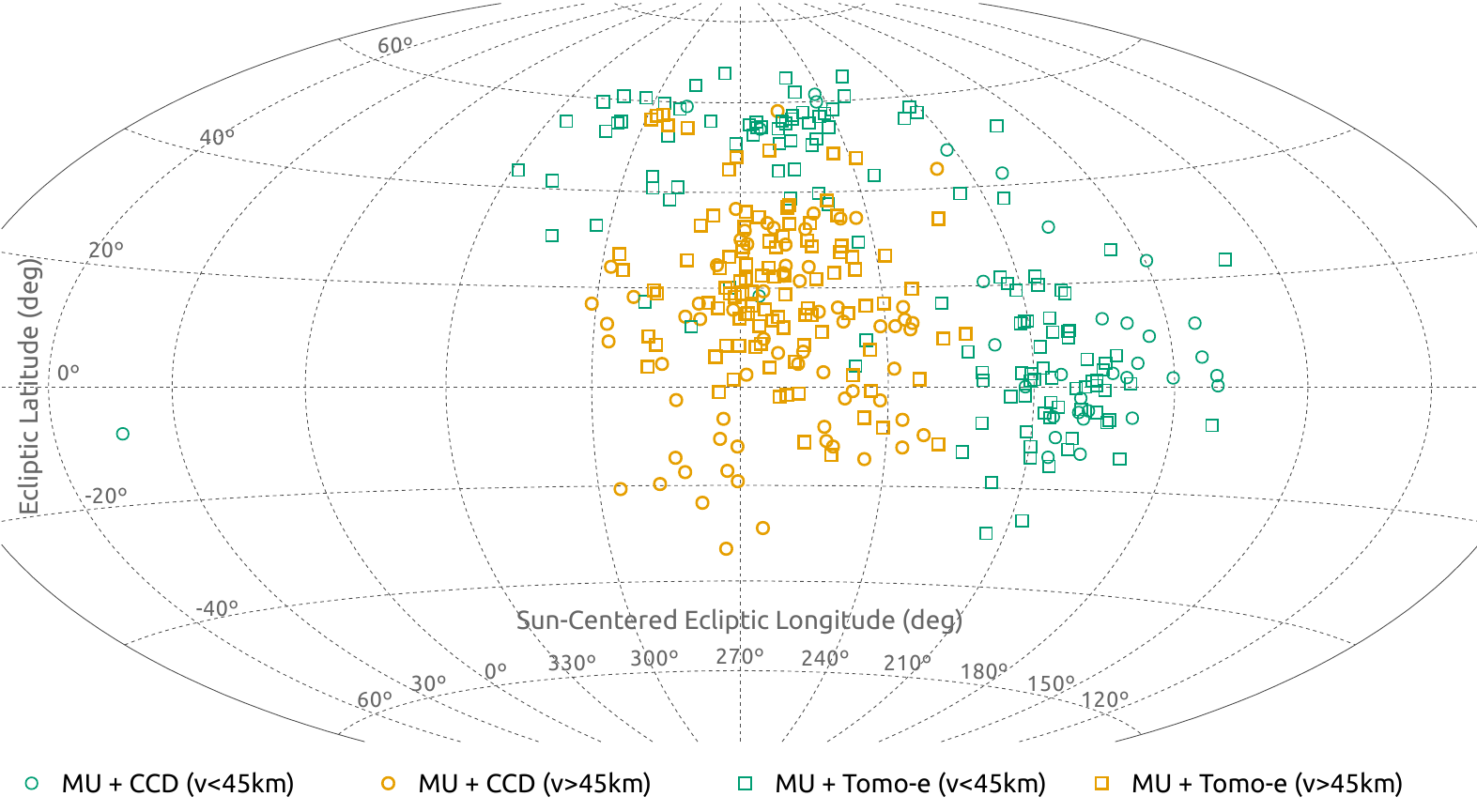}
\caption[The radiant distribution of the simultaneous meteors.]{The radiant distribution of the simultaneous meteors in the Sun-centered Ecliptic coordinates. The data of \runA are shown by the circle symbols, while those of \runB are shown by the square symbols. The meteors whose geocentric velocities are faster than $45\,\mathrm{km\,s^{-1}}$ are shown in orange, while the meteors slower than $45\,\mathrm{km\,s^{-1}}$ are shown in green.}
\label{fig:results:radiant}
\end{figure*}

Figure~\ref{fig:results:radiant} shows the distribution of the meteor radiants in the Hammer projection of the ecliptic latitude and Sun-centered ecliptic longitude coordinates. The meteors whose geocentric velocity is faster than $45\,\mathrm{km\,s^{-1}}$ are shown by the orange symbols. The distribution of the faster population is consistent with the apex sources \citep[e.g,][]{hawkins_radio_1956}, while the slower populations are possibly attributed to the antihelion and north toroidal sources \citep[e.g.,][]{hawkins_radio_1956,stohl_seasonal_1968}. No significant concentration is found in Figure~\ref{fig:results:radiant}, suggesting that possible contributions from meteor showers were successfully removed.

\begin{figure*}
\centering
\includegraphics[width=0.9\linewidth]{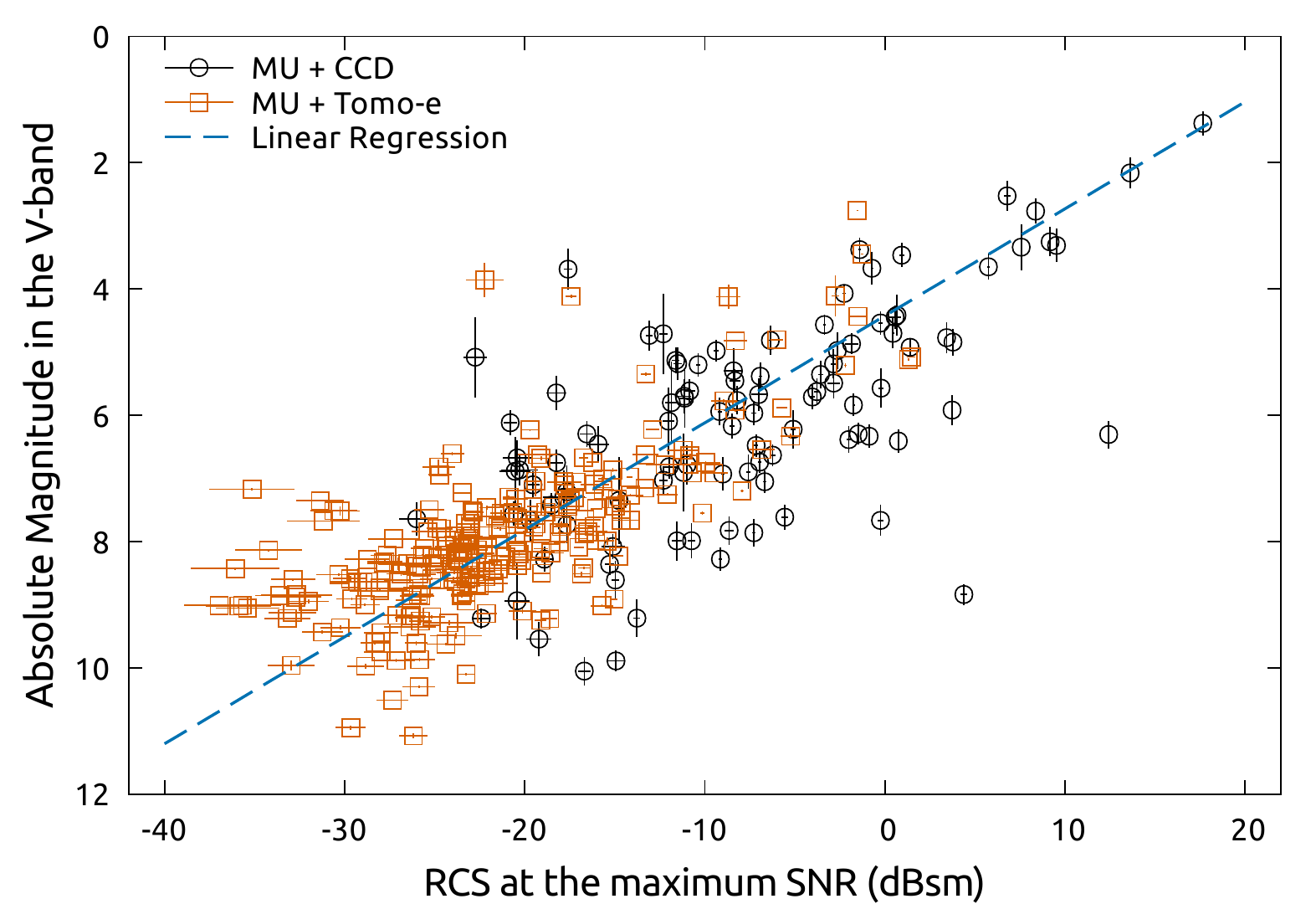}
\caption[The relationship between the RCS and $V$ magnitude.]{The relationship between the radar cross section and the optical absolute magnitude in the $V$-band. The meteors of \runA and \runB are shown by the black circle and red square symbols, respectively. The blue dashed line indicates the linear regression line.}
\label{fig:results:rcsmag}
\end{figure*}

Figure~\ref{fig:results:rcsmag} shows the optical absolute magnitude against the RCS, \revision{0206}{which is the value at the maximum signal-to-noise ratio (SNR) along the whole radar trajectory. On the other hand, the brightest magnitudes were independently measured in the optical observations. Thus, the section where the RCS was measured and the section where the optical brightness was measured could be different. A possible uncertainty due to this time difference is discussed later.} The data of \runA and \runB seem to follow the same trend, where the optical magnitude became brighter as the RCS bacame larger. The trend was approximated by a linear regression line. Since there was large scatter in Figure \ref{fig:results:rcsmag}, trial regression lines were calculated both in terms of the RCS and the magnitude via a least-square method and then the line with the averaged slope was adopted as the representative regression line:
\begin{equation}
\label{eq:results:rcsmag}
M_V = -(0.169{\pm}0.006){\times}A + (4.43{\pm}0.13),
\end{equation}
where $M_V$ is the meteor absolute magnitude in the $V$-band and $A$ is the radar cross section in units of dBsm (decibel relative to $1\,\mathrm{m^2}$). The representative regression line is shown by the blue dashed line in Figure~\ref{fig:results:rcsmag}. \revision{0206}{The uncertainties were estimated by the bootstrapping method. It should be noted that, since the color-term correction was not applied, the present result could suffer from a systematic bias due to different spectral responses of the cameras. Such a possible bias was not taken into account in the uncertainties in Equation~(\ref{eq:results:rcsmag}).}

\section{Discussion}
\label{sec:discussion}
\subsection{The relationship between the RCS and the optical magnitude}
Figure~\ref{fig:results:rcsmag} indicates that the data follow a single and linear relationship. No apparent deviation from the regression line is confirmed. The dependence of the relationship on the meteor speed is investigated by splitting the sample into the fast (${>}\,45\,\mathrm{km\,s^{-1}}$) and slow (${\leq}\,45\,\mathrm{km\,s^{-1}}$) members, but no significant difference is confirmed. This suggests that sporadic meteors from the apex, antihelion, and north toroidal sources follow the same relationship.

The scatter in Figure~\ref{fig:results:rcsmag} is much larger than the errors of the data. \revision{0206}{This may in part be due to the fact that the sections where the RCS and the optical magnitude were measured were close but not exactly the same. Both the RCS and the optical brightness are variable along the trajectory \citep[e.g., \textit{see} Figure~11 in][]{brown_simultaneous_2017}. The variation in the RCS and the optical brightness should contribute to the scatter. Especially, fragmentation may cause a sudden increase in the RCS and optical brightness, but the impacts of the fragmentation on the RCS and the magnitude are not necessarily the same \citep{campbell-brown_high-resolution_2013,brown_simultaneous_2017}. Fragmentation may cause pulsations in head echo RCS curves due to interference from two or more scattering centers \citep{kero_three-dimensional_2008}, while the luminosity produced is assumed to be proportional to the total kinetic energy lost by the meteoroid \citep{campbell-brown_high-resolution_2013}, thus proportional to the increased cross-sectional area to mass of the fragments. \citet{weryk_canadian_2013} reported that 17\% of meteors detected by the CAMO system showed clear signs of fragmentation. Similarly, a significant fraction of the data in Figure~\ref{fig:results:rcsmag} could be affected by fragmentation. This may partly explain the large scatter: ${\sim}2\,\mathrm{mag}$ in the optical magnitude and ${\sim}10\,\mathrm{dBsm}$ in the RCS, which are roughly consistent with the changes caused by the fragmentation.} Several studies have suggested that the ratio of the emission to ionization coefficients depends on the velocity \citep{saidov_luminous_1989,jones_theoretical_1997,weryk_simultaneous_2013}. This dependence may contribute to the scatter, but the difference in the distributions was not confirmed in the present data. A variety of chemical compositions may contribute to the scatter. Spectroscopic observations are required to confirm it.

Although the origin of the large scatter has not been identified, we tentatively conclude that the relationship between the RCS and the optical magnitude is well approximated by a linear function over a magnitude range of about $1$--$9\,\mathrm{mag}$, but we do not exclude the possibility that any deviations from the linear relationship are hidden within the scatter. \revision{0206}{The possibility that the relationship is variable or there are multiple relationships is not excluded as well. The scatter can be attributed to possible variations in the relationship. In such a case, the present relationship is considered to be an averaged one over the current dataset. Note that the uncertainties in Equation~(\ref{eq:results:rcsmag}) were derived with the assumption that the relationship is unique among the dataset, where the variation in the relationship was not taken into account.}

\revision{0206}{A scattering model of a head echo plasma was developed by \citet{close_technique_2004}, which provided a method to estimate the meteoroid mass, referred as to \textit{scattering mass}, from the head echo RCS \citep{close_new_2005}. \citet{close_meteor_2007} investigated the dependence of the head echo RCS on the electron line density ($q$), the scattering mass ($m_\mathrm{s}$), the velocity ($v$), and the mean-free-path ($l$). A multivariate regression provided a relationship $q \propto m^{1.08} v^{3.09} l^{-0.11}$ and the head echo RCS ($A$) was approximated as $A \propto q^{1.05} \propto m_\mathrm{s}^{1.05{\times}1.08}$. Here, we simply assume that the brightness of the meteor approximately follows $I \propto m_\mathrm{p}$, where $m_\mathrm{p}$ is the photometric mass. By equating $m_\mathrm{s}$ with $m_\mathrm{p}$, a relationship between the magnitude ($M_V$) and the head echo RCS is derived: $M_{V} = -0.22A+\text{const.}$ The slope of this relationship is similar to but steeper than the present result. This may implicate that the dependence of the head echo RCS is larger than suggested in \citet{close_meteor_2007}, or that the dependence of the optical brightness on the meteoroid mass is smaller. Note that the comparison above is highly simplified. A model calculation which evaluates the head echo RCS and the optical brightness simultaneously is required.}

\citet{nishimura_high_2001} provided a conversion function from the optical magnitude to the radar received power in units of $\mathrm{dB}$, where the optical magnitude decreased by about $0.3\,\mathrm{mag}$ when the radar power increased by $1\,\mathrm{dB}$, corresponding to the slope of $-0.3$. Since their result was derived from 20 simultaneous meteors, the slope could be affected by a considerable uncertainty. We estimate the uncertainty of the slope in case that the sample size is limited to 20 using a bootstrap sampling method: 20 meteors are randomly selected from the present 332 simultaneous meteors and a slope is derived by a least-square method. This process is repeated 1,000 times. Then, the posterior probability distribution of the slope is approximated by the distribution of the derived slopes. The 95\% confidence interval is $(-0.13, -0.32)$. Thus, we presume that the present slope is marginally consistent with that in \citet{nishimura_high_2001}. \citet{brown_simultaneous_2017} presented the relationship between the RCS the optical magnitude based on their 105 simultaneous meteors (\textit{see}, Figure~9). While no regression line was presented, the optical magnitude decreased roughly by $0.05$--$0.16\,\mathrm{mag}$ when the RCS increased by $1\,\mathrm{dB}$. The slope seems smaller than that of the present result. The optical magnitudes ranged over roughly $0$--$6\,\mathrm{mag}$ in \citet{brown_simultaneous_2017}, while only a handful of meteors were detected in this magnitude range in the present work. The possibility that the relationship changes around $5\,\mathrm{mag}$ is not excluded.

The data of \runA and \runB were obtained using different systems and on different days. Equation~(\ref{eq:results:rcsmag}) could suffer from systematic errors, such as the difference in spectral response and the annual variations. Further observations are required to evaluate the uncertainty of Equation~(\ref{eq:results:rcsmag}). The data of \runB were obtained in only four nights. This suggests that the combination of the MU radar and Tomo-e Gozen is promising to investigate the annual and diurnal variations in the relationship between the RCS and the optical magnitude.

\subsection{Meteor Luminosity Function of the MU radar Meteor Head Echo Database}
A luminosity function of visible meteors has been widely approximated by \revision{0206}{a power-law} function \citep{hawkins_influx_1958}:
\begin{equation}
\label{eq:discussion:lf}
\log_{10}N({<}M) = \log_{10}N_0 + M\log_{10}r,
\end{equation}
where $N({<}M)$ and $N_0$ are the event rates of meteors brighter than $M$-th and zero-th magnitudes, respectively, and $r$ defines the slope of the distribution, generally referred as to \textit{the population index}. \citet{cook_flux_1980} suggested that the luminosity function was well approximated by Equation~(\ref{eq:discussion:lf}) from $-2.4$ to $12\,\mathrm{mag}$. Here, Equation~(\ref{eq:results:rcsmag}) is applied to the data collected with the MU radar from 2009 to 2015, and the luminosity function of the meteors detected by the MU radar is investigated. The data consist of 157043 meteors in total. In the following discussion, we assume that the contributions from meteor showers are negligible.

\begin{figure*}
\centering
\includegraphics[width=0.495\linewidth]{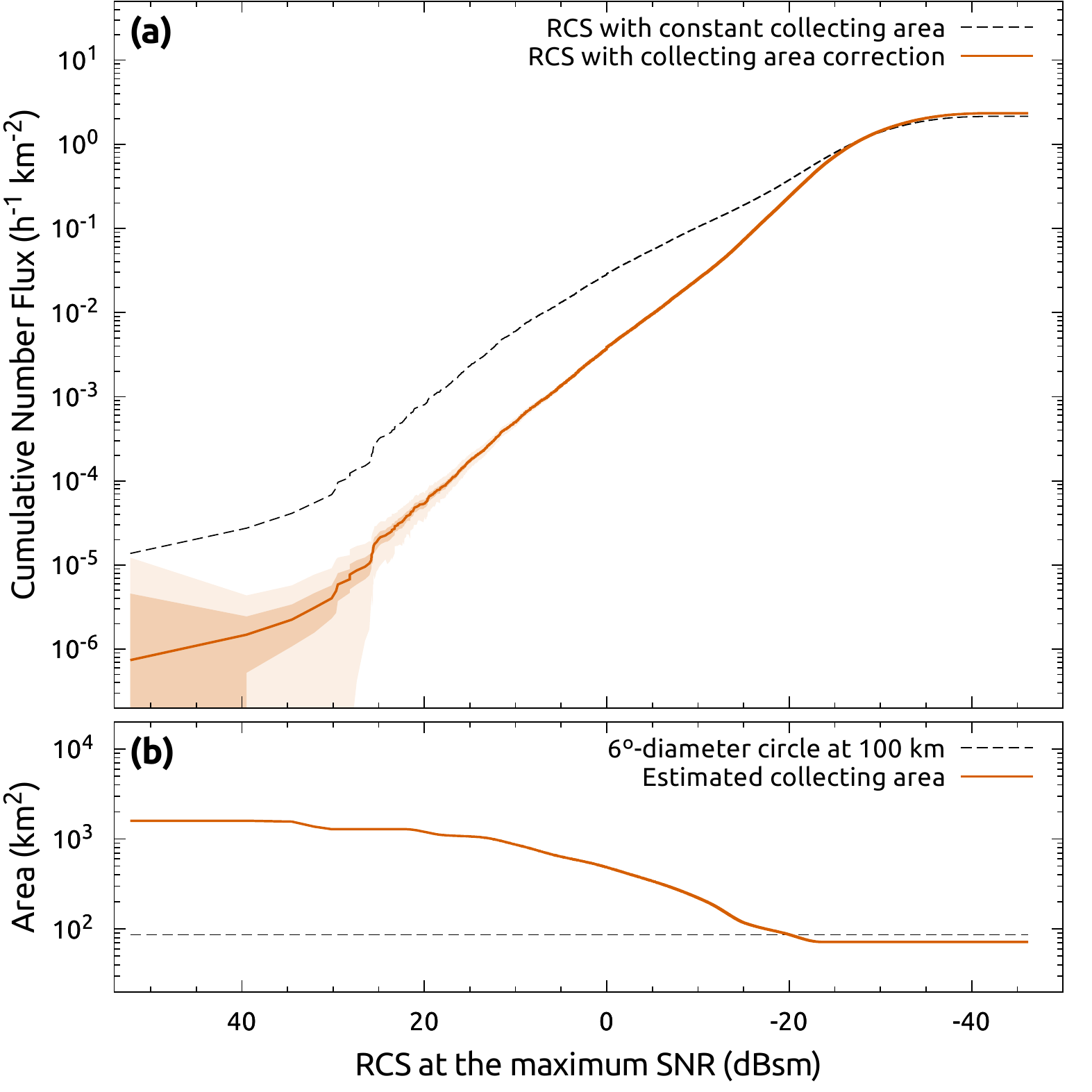}
\includegraphics[width=0.495\linewidth]{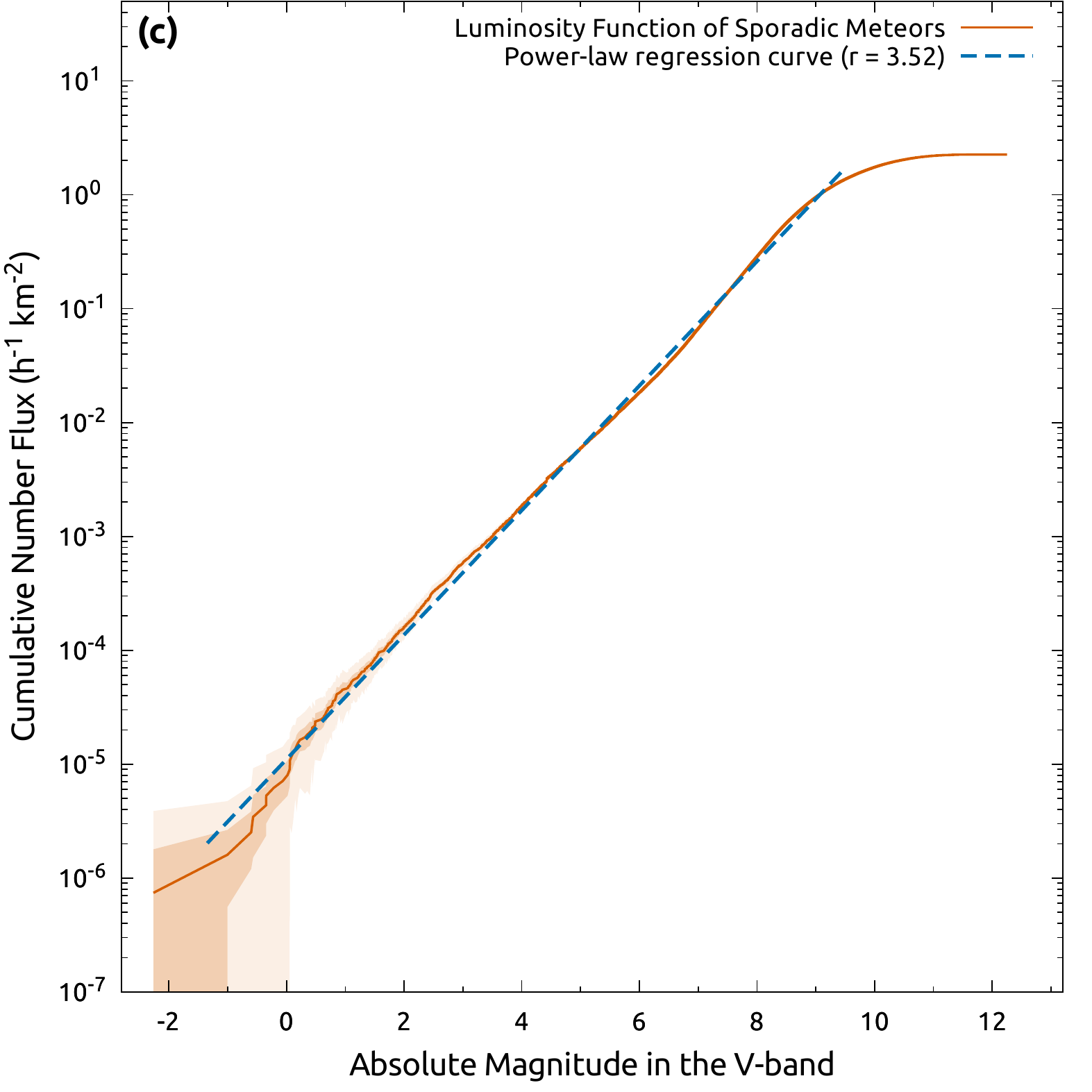}
\caption{The luminosity function of sporadic meteors. Panel (a) shows the cumulative number flux against the RCS. The black dashed line indicates the number flux assuming that the meteor collecting area does not depend on the RCS. The number flux shown by the red line properly takes into account the dependency. The shaded regions indicate $1\sigma$- and $3\sigma$-level uncertainties. The dependency of the meteor collecting area on the RCS is illustrated in Panel (b) in the red line. The black dashed line is the constant collecting area for reference. Panel (c) shows the luminosity function of sporadic meteors. The uncertainties are the same as in Panel (a). A regression curve is shown by the blue dashed line.}
\label{fig:discussion:luminosity}
\end{figure*}

Panel (a) of Figure~\ref{fig:discussion:luminosity} illustrates the cumulative number flux against the RCS in units of $\mathrm{h^{-1}\,km^{-2}}$. The black dashed line is the distribution assuming that the meteor collecting area is uniformly $86.3\,\mathrm{km^2}$ (a disk of $6\degree$ in diameter at $100\,\mathrm{km}$ in altitude). The meteor collecting area of the MU radar, however, depends on the RCS, since bright meteors can be detected in the side lobes of the beam \citep{kero_first_2011}. The dependence of the collecting area on the RCS is derived as follows; The distance from the beam center at $100\,\mathrm{km}$ ($R_{100\,\mathrm{km}}$) is derived for each meteor by interpolating or extrapolating the trajectory. The area of ${\pi}R_{100\,\mathrm{km}}^2$ is calculated. The data are divided into groups in the RCS and the median of the area among each group is adopted as the collecting area as a function of the RCS. The derived collecting area against the RCS is shown in Panel (b) of Figure~\ref{fig:discussion:luminosity}. The cumulative number flux calculated based on the derived collecting area is shown by the red line with the $1\sigma$- and $3\sigma$-uncertainty regions in Figure~\ref{fig:discussion:luminosity}. The cumulative number flux larger than $25\,\mathrm{dBsm}$ is highly uncertain. The cumulative number flux peaks out around ${-}25\,\mathrm{dBsm}$ simply due to the detection limit. In a range from $-20$ to $20\,\mathrm{dBsm}$, the cumulative number flux seems well approximated by a linear function.

The cumulative number flux against the RCS is converted into the luminosity function by applying Equation~(\ref{eq:results:rcsmag}). Panel (c) of Figure~\ref{fig:discussion:luminosity} illustrates the luminosity function in the red line with the $1\sigma$- and $3\sigma$-uncertainty regions. The luminosity function basically follows an exponential law, which is consistent with previous works \citep[e.g.,][]{hawkins_influx_1958,hawkes_television_1975,cook_flux_1980,ohsawa_luminosity_2019}. The detection limit corresponds to about $10\,\mathrm{mag}$, which is roughly consistent with the detection limit for EISCAT in \citet{pellinen-wannberg_meteor_1998}. The population index is derived as $r = 3.52{\pm}0.12$ by fitting a linear function between $-1.5$ and $9.5\,\mathrm{mag}$, taking into account the systematic uncertainty from Equation~(\ref{eq:results:rcsmag}).

\begin{table*}
\centering
\caption{The population indexes in literature}
\label{tab:discussion:index}
\begin{tabular}{llll}
\toprule
\multicolumn{1}{c}{Reference}
& \multicolumn{1}{c}{$r$-index}
& \multicolumn{1}{c}{$s$-index}
& \multicolumn{1}{c}{Comment} \\\midrule
\citet{hawkins_influx_1958}
& ${\sim}3.45$
& ${\sim}2.34$
& $-2$--$3\,\mathrm{mag}$, photographic \\
\citet{kresakova_magnitude_1966}
& $\phantom{\sim}3.5$
& $\phantom{\sim}2.35$
& $-4$--$6\,\mathrm{mag}$, 21996 visual meteors \\
\citet{hughes_diurnal_1972}
& \multicolumn{1}{c}{---}
& $\phantom{\sim}2.04{\pm}0.04$
& 30000 HF radar echoes \\
\citet{clifton_television_1973}
& ${\sim}3.17$
& ${\sim}2.252$
& $7$--$11\,\mathrm{mag}$, TV observation\\
\citet{hughes_influx_1974}
& $\phantom{\sim}3.73{\pm}0.07$
& $\phantom{\sim}2.43{\pm}0.02$
& $-6$--$0\,\mathrm{mag}$, 10287 visual meteors \\
\citet{hawkes_television_1975}
& \multicolumn{1}{c}{---}
& $\phantom{\sim}2.02{\pm}0.04$
& $3$--$7\,\mathrm{mag}$, TV observation \\
\citet{stohl_magnitude_1976}
& $\phantom{\sim}3.70$
& \multicolumn{1}{c}{---}
& 12867 visual meteors \\
\citet{cook_flux_1980}
& $\phantom{\sim}3.41$
& $\phantom{\sim}2.335$
& $7$--$12\,\mathrm{mag}$, phototubes \\
\citet{rendtel_population_2004}
& $\phantom{\sim}2.95{\pm}0.06$
& $\phantom{\sim}2.17{\pm}0.03$
& 301499 visual meteors, IMO VMDB \\
\citet{ohsawa_luminosity_2019}
& $\phantom{\sim}3.1{\pm}0.4$
& \multicolumn{1}{c}{---}
& $3$--$9\,\mathrm{mag}$, video-rate magnitude \\
The present result
& $\phantom{\sim}3.52{\pm}0.12$
& $\phantom{\sim}2.46{\pm}0.09$
& $0$--$9\,\mathrm{mag}$, MURMHED \\
\bottomrule
\end{tabular}
\end{table*}

\begin{figure*}
\centering
\begin{minipage}[b]{0.49\linewidth}
\includegraphics[width=\linewidth]{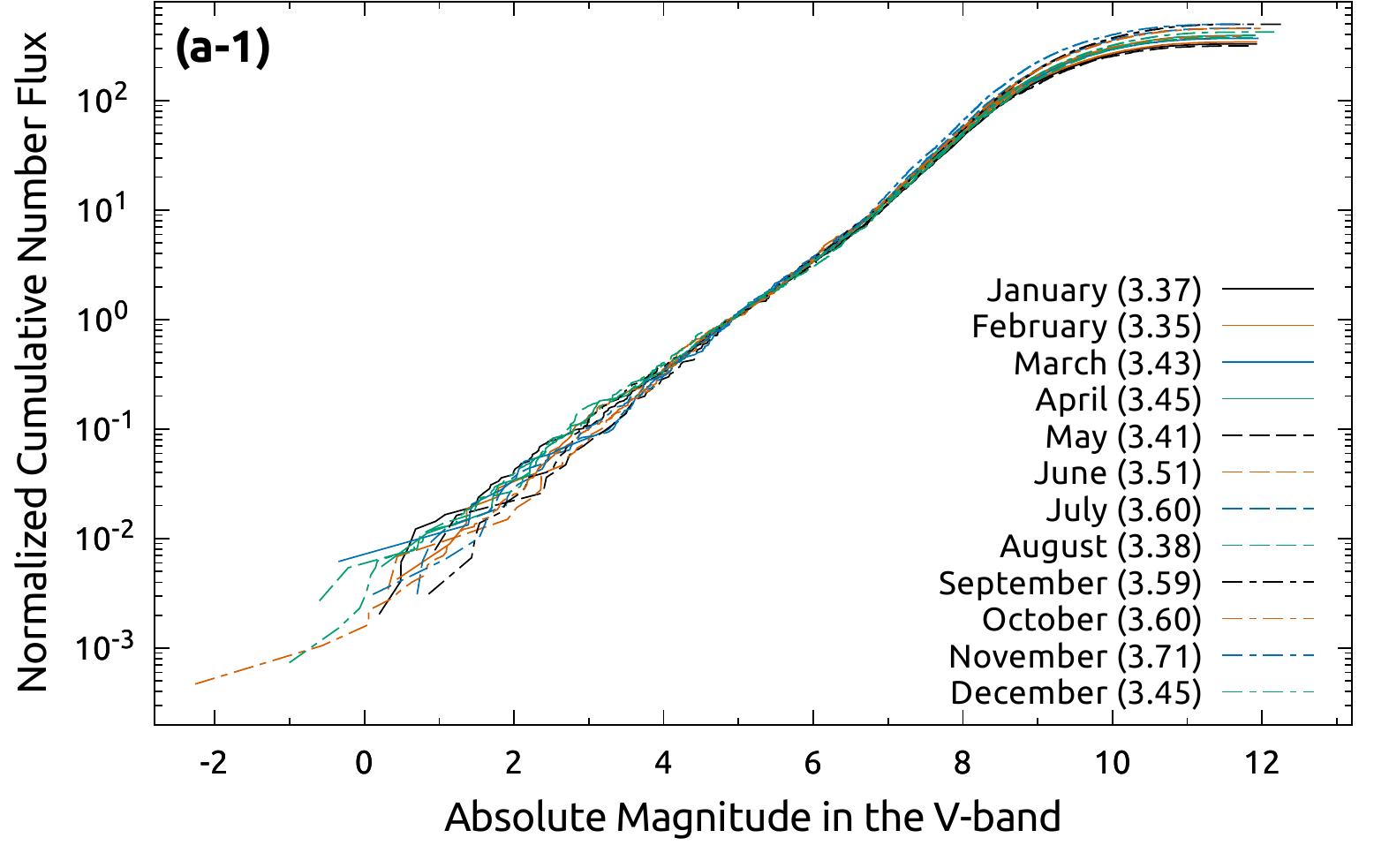}
\includegraphics[width=\linewidth]{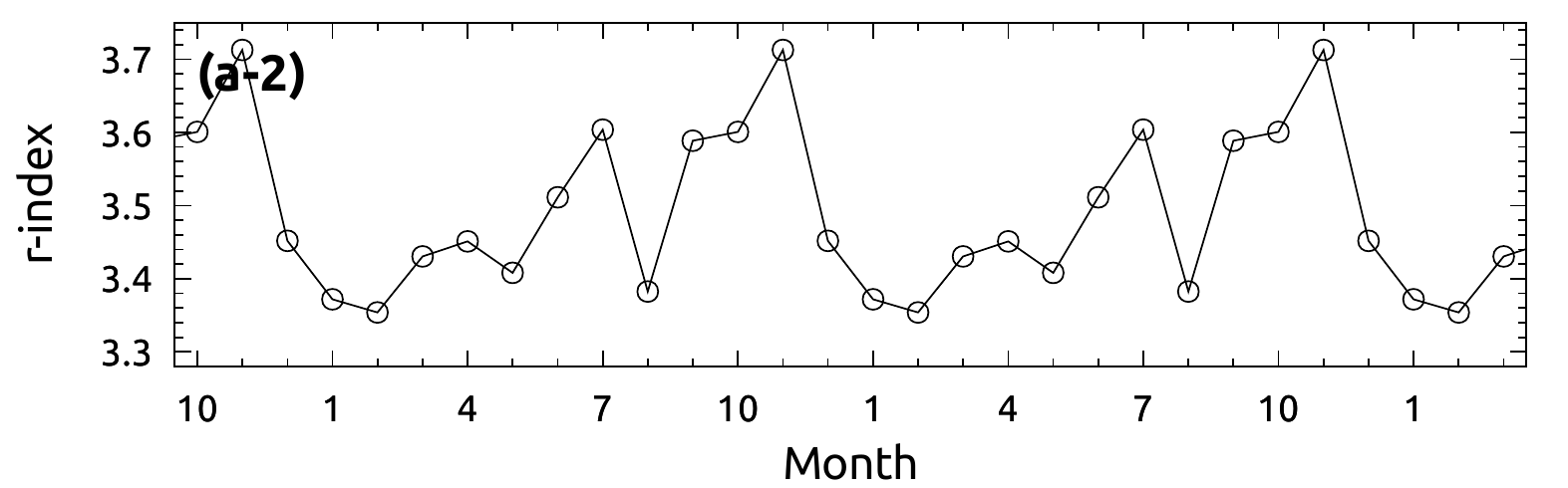}
\end{minipage}
\hspace{0.5em}%
\begin{minipage}[b]{0.49\linewidth}
\includegraphics[width=\linewidth]{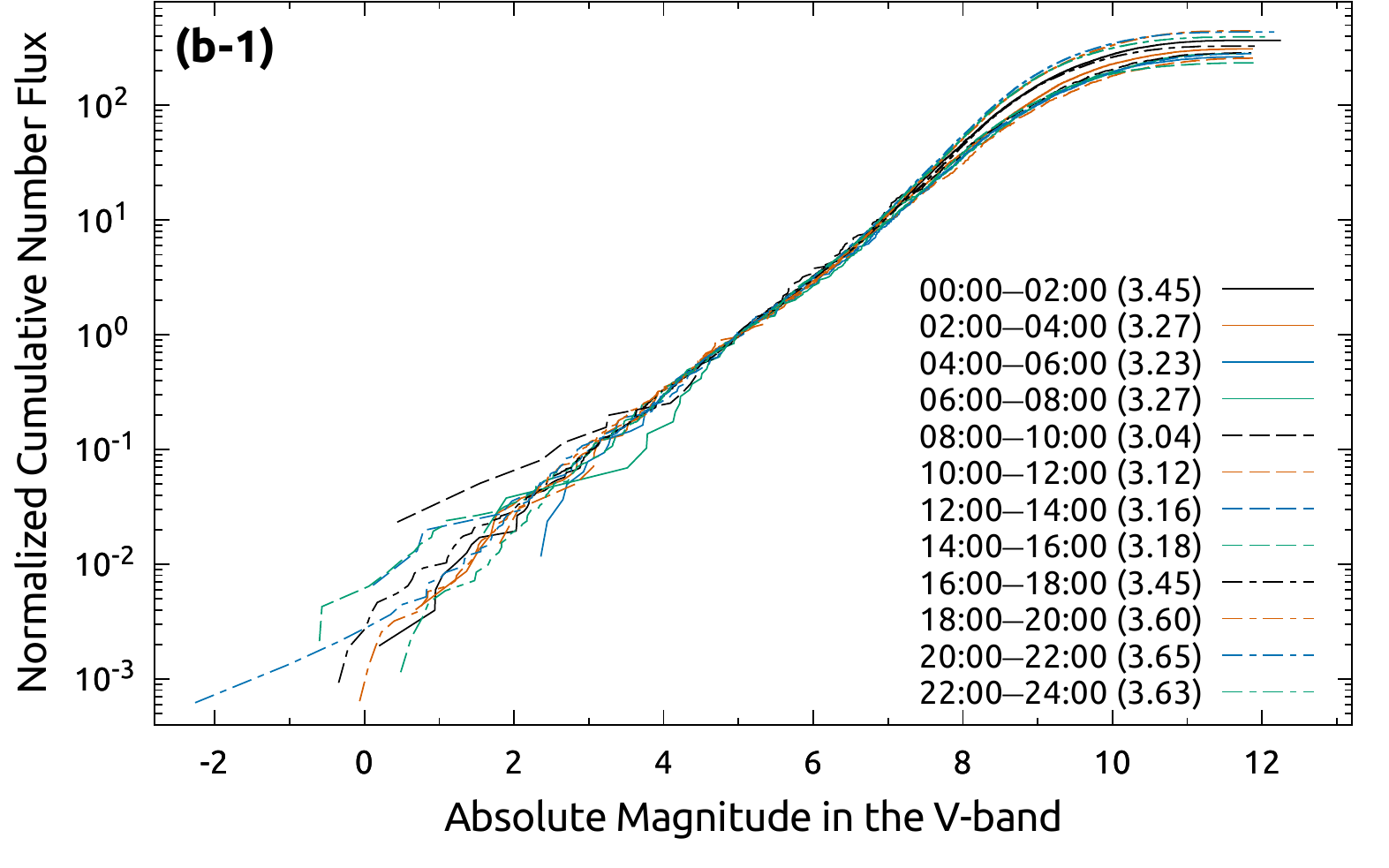}
\includegraphics[width=\linewidth]{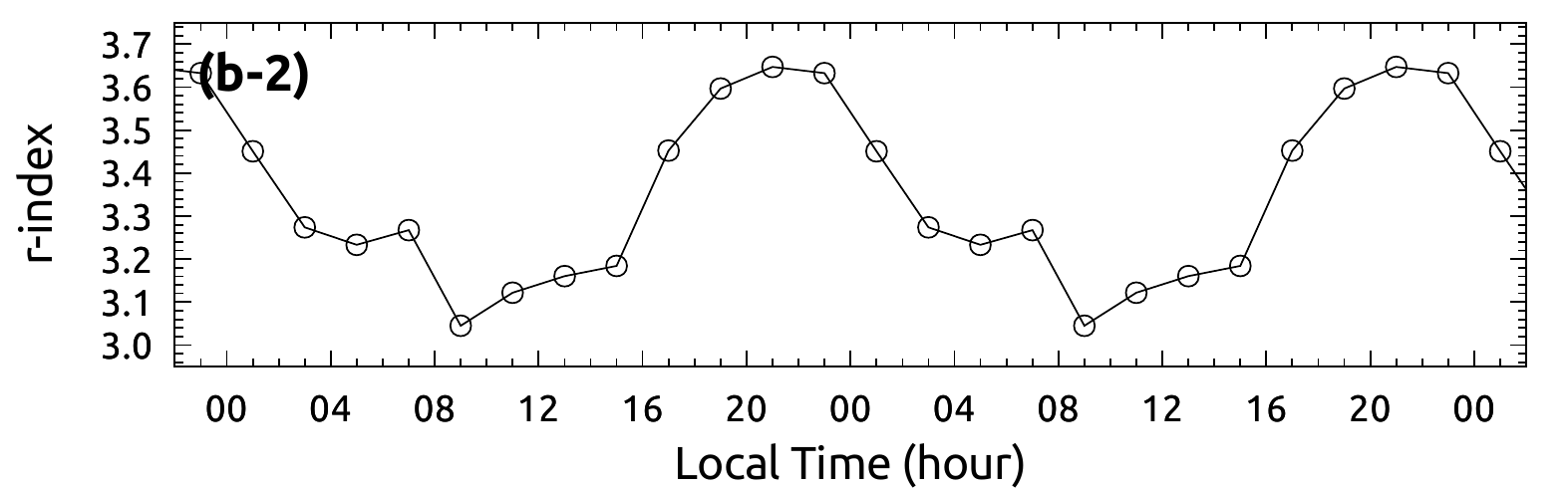}
\end{minipage}
\caption{Panel (a-1) shows the month-by-month variation in the luminosity functions, while the diurnal variation in the luminosity functions is shown in Panel (b-1). All the luminosity functions are normalized around $5\,\mathrm{mag}$. The numbers in the legends of each panel indicate the population indexes. The annual and diurnal variations in the population index are shown in Panels (a-2) and (b-2), respectively. The times are based on the detection time stamps in the local time (Japan Standard Time).}
\label{fig:discussion:monthly-and-hourly}
\end{figure*}

Table~\ref{tab:discussion:index} lists some population indexes reported in literature. A more detailed overview found in \citet{rendtel_population_2004}. The population indexes have been measured in different methods and different magnitude ranges. As shown in Table~\ref{tab:discussion:index}, there is a considerable variation in the reported population indexes. The present result is roughly consistent with \citet{hawkins_variation_1956}, \citet{kresakova_magnitude_1966}, \citet{cook_flux_1980}, and \citet{ohsawa_luminosity_2019}.

Several studies have pointed out the annual and diurnal variations in the population index \citep[e.g.,][]{hughes_diurnal_1972,rendtel_population_2004}. Panel (a-1) of Figure~\ref{fig:discussion:monthly-and-hourly} illustrates the annual variation in the luminosity function. The luminosity functions are normalized around $5\,\mathrm{mag}$. The annual average of the population index is about 3.49 with a standard deviation of 0.11. Panel (a-2) shows a possible annual variation in the population index: it tends to be high in \revision{0206}{July--September} and low in \revision{0206}{December--March}. \citet{rendtel_population_2004}, however, suggested that the population index increases (decreases) in winter (summer) in the northern hemisphere, which is the opposite to the present result. A similar trend was also reported by \citet{stohl_magnitude_1976}. \citet{rendtel_population_2004} complied the dataset from observations in 1998--2003 and the observations in \citet{stohl_magnitude_1976} were in 1944--1950. The present data were collected 2009--2015. This could be explained in case that the trend in the population index changed in the timescale of decades. Limiting magnitudes in \citet{stohl_magnitude_1976} and \citet{rendtel_population_2004} were about $4$ and $6\,\mathrm{mag}$, respectively. Instead, the present result is biased toward meteors in a magnitude range of $4$--$8\,\mathrm{mag}$. This humbly implies that brighter and fainter parts of the luminosity function show different annual variations.

Panel (b-1) of Figure~\ref{fig:discussion:monthly-and-hourly} shows the diurnal variation in the luminosity function, normalized as in Panel (a-1). All the times are in the local time (Japan Standard Time). Panel (b-2) shows a diurnal variation in the population index. The averaged population index is 3.33 with a standard deviation of 0.21, suggesting the diurnal variation is more significant than the annual variation. The population index almost constantly decreases \revision{0206}{from 21:00 and reaches a minimum around 09:00}. This trend is roughly consistent with \citet{hughes_diurnal_1972} and \citet{rendtel_population_2004}, although \citet{stohl_magnitude_1976} reported an almost constant population index in this period. The present result suggests that the diurnal variation in the luminosity function does not depend on the optical magnitude, or the meteoroid mass.

Note that the present population index is heavily dependent on Equation~(\ref{eq:results:rcsmag}). In the discussion above, we naively assume that the relationship between the RCS and the magnitude is constant. The results could be affected by any possible annual or diurnal variations in the relationship. \revision{0206}{Figure~\ref{fig:results:radiant} indicates that a large part of the meteors simultaneously detected by radar and optically were attributed to the antihelion and north toroidal sources. The meteors from these sources are typically slower than those from the apex sources, which are dominant in the radar observation \citep{kero_2009-2010_2012}. Thus, the meteors which were used to derive Equation~(\ref{eq:results:rcsmag}) could be biased toward a larger part of the meteor population. Part of the diurnal variation can be attributable to such an observational bias.} More systematic observations are required to examine the validity of the present results.

\begin{figure*}
\centering
\includegraphics[width=0.9\linewidth]{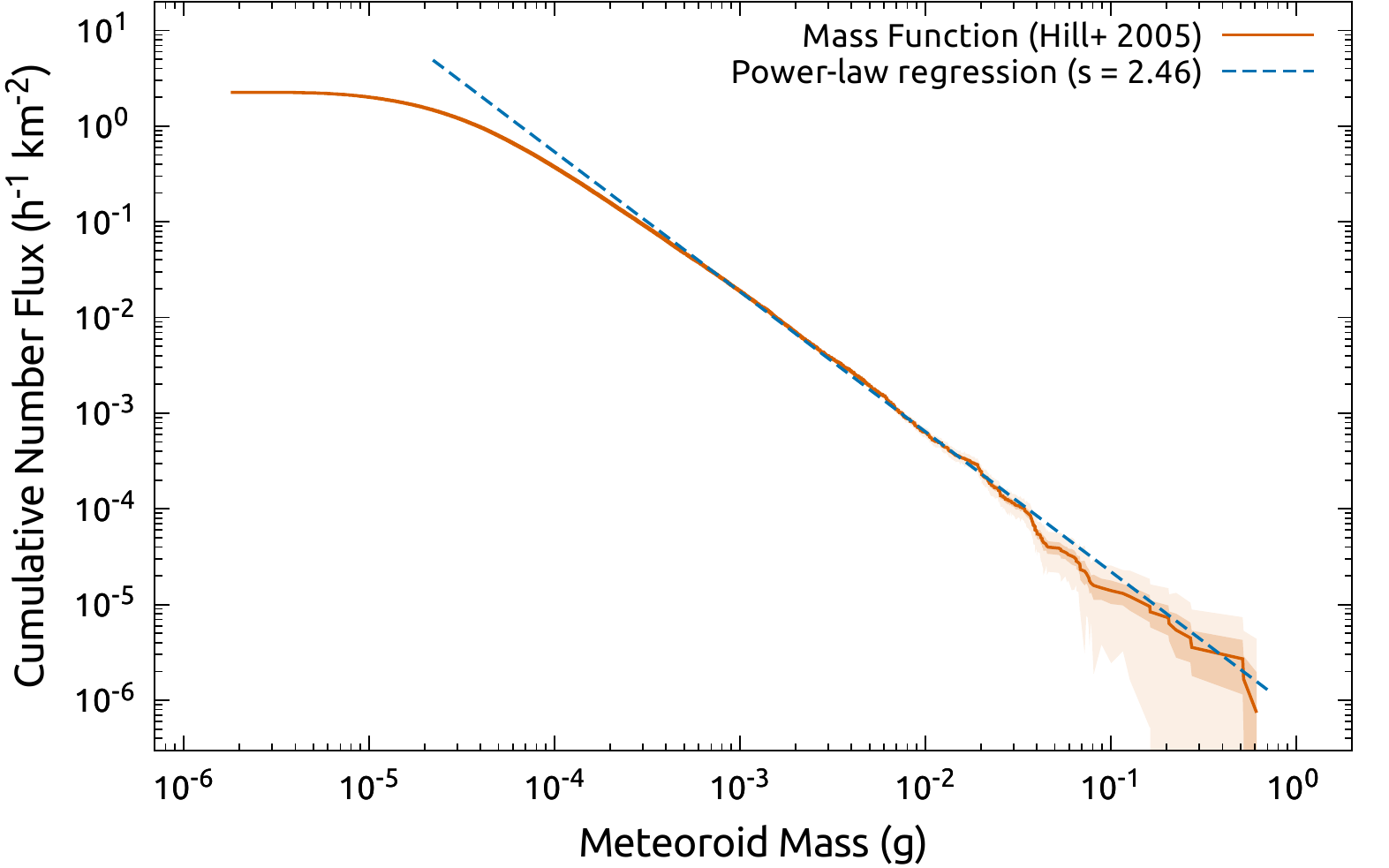}
\caption{The mass function of sporadic meteors is shown by the red line. The $1\sigma$- and $3\sigma$-level uncertainties are shown by the shaded regions. The blue dashed line indicates the regression curve.}
\label{fig:discussion:mass}
\end{figure*}

\revision{0206}{The optical magnitude is converted into mass using the ablation equations \citep[e.g.,][]{hawkes_quantitative_1975,ceplecha_meteor_1998}. When a meteoroid enters the atmosphere, it is decelerated and heated through interactions with atmospheric constitutions. The meteoroid loses its kinetic energy by decreasing bot its mass and velocity. Part of the lost energy is converted into light and observed in optical. The fraction of the meteoroid's kinetic energy which is converted into light is called the luminous efficiency. The thermal ablation of a meteoroid is calculated following the procedure given by \citet{hill_high_2005}. The parameters in the ablation equations are adopted from \citet{hill_high_2005}. A typical molecular mass and bulk density are respectively set to $50\,amu$ and $3300\,\mathrm{kg\,m^{-3}}$. The calculation ends when the meteoroid loses 99.999\% of the initial mass. The brightest magnitudes are calculated using the normal luminous equation for different meteoroid masses ($10^{1},\,10^{0},\,\ldots,\,10^{-7}\,\mathrm{g}$), radiant zenith angles ($0,\,30\degree,\,45\degree,\,60\degree$, and $90\degree$), and initial velocities ($10,\,20,\,\ldots,\,70\,\mathrm{km/s}$). The luminous efficiency is taken from \citet{hill_high_2005}, but scaled by 0.081\footnote{The scaling factor in \citet{weryk_simultaneous_2013} is $0.453{\times}0.28$, which was optimized to the observation in the \textit{R}-band. It is further multiplied by 0.638, which is the band width ration of \textit{V}-band to the \textit{R}-band.} as described in \citet{weryk_simultaneous_2013}. Then, a relationship to calculate the meteoroid mass from the magnitude, velocity, and radiant zenith angle is approximated by the following function:}
\begin{equation}
\label{eq:discussion:mass:h05}
\revision{0206}{
\log_{10}m = 2.76-0.38M_V-2.31\log_{10}V_{\infty}-1.07\log_{10}\cos{z},
}
\end{equation}
\revision{0206}{where $m$ is the mass in units of $\mathrm{g}$, $M_V$ is the optical magnitude, $V_\infty$ is the incident velocity in units of $\mathrm{km\,s^{-1}}$, and $z$ is the zenith angle of the radiant point. $V_\infty$ and $z$ are calculated from the trajectories measured by the MU radar. The deviation from Equation~(\ref{eq:discussion:mass:h05}) is typically up to about $0.2\,\mathrm{dex}$. The derived mass depends on the selection of a luminous coefficient. While the current calculation uses the scaled luminous coefficient of \citet{hill_high_2005}, larger luminous coefficients were reported by \citet{ceplecha_fragmentation_2005} and \citet{weryk_simultaneous_2013}. If we adopt those luminous coefficients, the derived mass could be smaller by about an order of magnitude.}

The cumulative mass distribution, or the mass function, is illustrated in Figure~\ref{fig:discussion:mass}. \revision{0206}{The mass function peaks out around $10^{-5}\,\mathrm{g}$, which is attributed to the mass detection limit.} The mass function is expected to follow a power-law function:
\begin{equation}
\label{eq:discussion:mass_index}
\log_{10}N({>}m) = \log_{10}N_1 - (s-1)\log_{10}m,
\end{equation}
where $m$ is the meteoroid mass in units of $g$, $N({>}m)$ and $N_1$ are the event rates of meteors larger than $m$ and $1\,\mathrm{g}$, respectively, and $s$ defines the slope of the mass function, referred as to \textit{the mass index}. \revision{0206}{The derived regression curve is shown by the blue line.} Some mass indexes in literature are listed in Table~\ref{tab:discussion:index}. The mass function deviates from the regression curve around $10^{-4}\,\mathrm{g}$. This deviation is likely not a real feature, but could be attributed to the observational bias: the detection limit is given by the RCS rather than the meteoroid mass. \revision{0206}{The derived mass index is $2.46{\pm}0.09$. This value is similar to that in \citet{hughes_influx_1974} and \citet{cook_flux_1980}. The mass flux to the Earth in the mass range of $10^{-5}$--$10^{0}\,\mathrm{g}$ is about $3{\times}10^3\,\mathrm{kg\,day^{-1}}$. \citet{ceplecha_influx_1992} provided an incremental mass flux of interplanetary bodies by compiling several observational researches. By integrating it from $10^{-5}$ to $10^{0}\,\mathrm{g}$, the mass flux was $1{\times}10^3\,\mathrm{kg\,day^{-1}}$, which is roughly coincident with our estimation. \citet{love_direct_1993}, however, estimated the mass flux in the similar mass range to be about $1{\times}10^5\,\mathrm{kg\,day^{-1}}$ from the examinations of impact craters on the Long Duration Exposure Facility satellite. The current estimate is more than an order smaller than their estimate. Note that the estimated mass flux depends on the selection of the luminous efficiency. The mass flux will be smaller if the luminous efficiencies of \citet{ceplecha_fragmentation_2005} and \citet{weryk_simultaneous_2013} are applied.}

\section{Conclusion}
\label{sec:conclusion}
We report the results of the two radar-and-optical simultaneous observations of faint meteors. The first observation run was in 2009--2010, using Middle and Upper Atmosphere Radar (MU radar) and an image-intensified CCD video camera equipped with a Canon $200\,\mathrm{mm}$ F/1.8 lens. In total, 103 simultaneous sporadic meteors were detected and a typical magnitude was about $6.2\,\mathrm{mag}$. The second observation run was carried out in 2018, using the MU radar and a mosaic CMOS camera, Tomo-e Gozen, installed in the 1.05-m Kiso Schmidt telescope. Consequently, 228 simultaneous sporadic meteors were detected. A typical magnitude was about $8.1\,\mathrm{mag}$. The velocity distribution of the detected meteors was bimodal. No apparent trend was found in the altitude and velocity of the meteor in terms of the optical brightness. The distribution of the radiant points indicates that the fast (${>}\,45\,\mathrm{km\,s^{-1}}$) and slow (${\leq}\,45\,\mathrm{km\,s^{-1}}$) populations are attributed to the apex sources and the antihelion or north-toroidal sources, respectively.

The optical magnitude calibrated to the $V$-band linearly decreased with increasing radar cross section (RCS) at the peak signal-to-noise ratio. The linear regression line was derived as: $M_V = (-0.169{\pm}0.006){\times}A + (4.43{\pm}0.13)$, where $M_V$ is the $V$-band magnitude and $A$ is the RCS in units of $\mathrm{dBsm}$. The slope of the regression line was marginally consistent with that in \citet{nishimura_high_2001}. The slope in \citet{brown_simultaneous_2017}, which was obtained in a magnitude range of $0$--$6\,\mathrm{mag}$, was smaller than the present result. Further observations are required to investigate the difference in the slopes. Although there is large scatter around the regression line, we confidently obtained a conversion function from the RCS to the optical magnitude.

By applying the derived conversion function to more than 150,000 meteors collected from the MU Radar Meteor Head Echo Database (MURMHED), we compile a luminosity function of faint sporadic meteors in units of $\mathrm{hour^{-1}\,km^{-2}}$. The luminosity function is well approximated by an exponential function in a magnitude range of $-1.5$--$9.5\,\mathrm{mag}$. The population index $r$ is constrained to be $3.52{\pm}0.12$ by fitting the exponential function. The luminosity function peaks out around $10\,\mathrm{mag}$, corresponding to the detection limit of the MU radar. The annual and diurnal variations in the luminosity function are investigated. The present result shows the opposite trend to the trends in literature \citep{stohl_magnitude_1976,rendtel_population_2004}. This is possibly explained by assuming the trend changes in a timescale of decades or that the trend depends on the optical magnitude. The present diurnal variation is consistent with previous works, suggesting that the diurnal variation is less dependent on the optical magnitude. The optical magnitude was converted into the mass of the meteoroid \revision{0206}{based on the thermal ablation theory and the luminous efficiency in \citet{hill_high_2005} but scaled by 0.081 as described in \citet{weryk_simultaneous_2013}. The size index $s$ is constrained to be $2.46{\pm}0.09$ by fitting a power-law function. The mass function peaks out around $10^{-5}\,\mathrm{g}$. We conclude that the MU radar is able to detect interplanetary particles of $10^{-5}$--$10^{0}\,\mathrm{g}$ in mass as meteors. The mass flux to the Earth in this mass range is estimated to be a few $10^3\,\mathrm{kg\,day^{-1}}$, but this amount could suffer from the uncertainty in the luminous efficiency.}

The present results are based on the relationship between the RCS and the optical magnitude. This is derived from the two observation runs, which were conducted with the different observing systems and in the different seasons. The derived relationship is possibly affected by systematic errors which are not taken into account in the current analysis. Further systematic observations are required to validate the present results. The combination of the MU radar and Tomo-e Gozen seems promising to conduct such observations and to investigate statistical characteristics of faint meteors.

\section*{Acknowledgments}
This research has been partly supported by Japan Society for the Promotion of Science (JSPS) Grants-in-Aid for Scientific Research (KAKENHI) Grant Numbers 26287106, 16H02158, 16H06341, 18H01272, 18H01261, 18H04575, and 18K13599. This research is also supported in part by the Japan Science and Technology (JST) Agency's Precursory Research for Embryonic Science and Technology (PRESTO), the Research Center for the Early Universe (RESCEU), of the School of Science at the University of Tokyo, and the Optical and Near-infrared Astronomy Inter-University Cooperation Program. The meteor head echo data (MURMHED) used in this study have been created by T. Nakamura (NIPR, Japan), J. Kero (IRF, Sweden) and members of the radar meteor head echo database group under the support by JSPS Grant-in-Aid for Publication of Scientific Research Results (KAKENHI Databases) Grant Number 258033. The MU radar belongs to and is operated by RISH (Research Institute of Sustainable Humanosphere), Kyoto University.


\begin{thebibliography}{79}
\expandafter\ifx\csname natexlab\endcsname\relax\def\natexlab#1{#1}\fi
\expandafter\ifx\csname url\endcsname\relax
  \def\url#1{\texttt{#1}}\fi
\expandafter\ifx\csname urlprefix\endcsname\relax\def\urlprefix{URL }\fi

\bibitem[{Andrei{\'c} et~al.(2014)Andrei{\'c}, Gural, {\v{S}}egon, Skoki{\'c},
  Korlevi{\'c}, Vida, Novoselnik, and Gostinski}]{andreic_results_2014}
Andrei{\'c}, {\v{Z}}., Gural, P., {\v{S}}egon, D., Skoki{\'c}, I.,
  Korlevi{\'c}, K., Vida, D., Novoselnik, F., Gostinski, D., 2014. Results of
  {CMN} 2013 search for new showers across {CMN} and {SonotaCo} databases
  {I}-0.5mm. WGN, Journal of the International Meteor Organization 42, 90--97.
\newline\urlprefix\url{http://adsabs.harvard.edu/abs/2014JIMO...42...90A}

\bibitem[{Andrei{\'c} et~al.(2013)Andrei{\'c}, {\v{S}}egon, Korlevi{\'c},
  Novoselnik, Vida, and Skoki{\'c}}]{andreic_ten_2013}
Andrei{\'c}, {\v{Z}}., {\v{S}}egon, D., Korlevi{\'c}, K., Novoselnik, F., Vida,
  D., Skoki{\'c}, I., 2013. Ten possible new showers from the {Croatian}
  {Meteor} {Network} and {SonotaCo} datasets. WGN, Journal of the International
  Meteor Organization 41, 103--108.
\newline\urlprefix\url{http://adsabs.harvard.edu/abs/2013JIMO...41..103A}

\bibitem[{Baldwin and Sheaffer(1971)}]{baldwin_ablation_1971}
Baldwin, B., Sheaffer, Y., 1971. Ablation and breakup of large meteoroids
  during atmospheric entry. Journal of Geophysical Research 76, 4653.
\newline\urlprefix\url{https://ui.adsabs.harvard.edu/abs/1971JGR....76.4653B/abstract}

\bibitem[{Boitnott and Savage(1972)}]{boitnott_light-emission_1972}
Boitnott, C.~A., Savage, H.~F., 1972. Light-{Emission} {Measurements} of {Iron}
  at {Simulated} {Meteor} {Conditions}. The Astrophysical Journal 174, 201.
\newline\urlprefix\url{http://adsabs.harvard.edu/abs/1972ApJ...174..201B}

\bibitem[{Brown et~al.(2017)Brown, Stober, Schult, Krzeminski, Cooke, and
  Chau}]{brown_simultaneous_2017}
Brown, P., Stober, G., Schult, C., Krzeminski, Z., Cooke, W., Chau, J.~L.,
  2017. Simultaneous optical and meteor head echo measurements using the
  {Middle} {Atmosphere} {Alomar} {Radar} {System} ({MAARSY}): {Data} collection
  and preliminary analysis. Planetary and Space Science 141, 25--34.
\newline\urlprefix\url{https://ui.adsabs.harvard.edu/abs/2017P%26SS..141...25B/abstract}

\bibitem[{Campbell-Brown et~al.(2013)Campbell-Brown, Borovi{\v{c}}ka, Brown,
  and Stokan}]{campbell-brown_high-resolution_2013}
Campbell-Brown, M.~D., Borovi{\v{c}}ka, J., Brown, P.~G., Stokan, E., 2013.
  High-resolution modelling of meteoroid ablation. Astronomy and Astrophysics
  557, A41.
\newline\urlprefix\url{http://adsabs.harvard.edu/abs/2013A%26A...557A..41C}

\bibitem[{Campbell-Brown et~al.(2012)Campbell-Brown, Kero, Szasz,
  Pellinen-Wannberg, and Weryk}]{campbell-brown_photometric_2012}
Campbell-Brown, M.~D., Kero, J., Szasz, C., Pellinen-Wannberg, A., Weryk,
  R.~J., 2012. Photometric and ionization masses of meteors with simultaneous
  {EISCAT} {UHF} radar and intensified video observations. Journal of
  Geophysical Research (Space Physics) 117~(A9), A09323.
\newline\urlprefix\url{https://ui.adsabs.harvard.edu/abs/2012JGRA..117.9323C/abstract}

\bibitem[{Ceplecha(1992)}]{ceplecha_influx_1992}
Ceplecha, Z., 1992. Influx of interplanetary bodies onto earth. Astronomy and
  Astrophysics 263, 361--366.
\newline\urlprefix\url{http://adsabs.harvard.edu/abs/1992A%26A...263..361C}

\bibitem[{Ceplecha et~al.(1998)Ceplecha, Borovi{\v{c}}ka, Elford, Revelle,
  Hawkes, Porub{\v{c}}an, and {\v{S}}imek}]{ceplecha_meteor_1998}
Ceplecha, Z., Borovi{\v{c}}ka, J., Elford, W.~G., Revelle, D.~O., Hawkes,
  R.~L., Porub{\v{c}}an, V., {\v{S}}imek, M., 1998. Meteor {Phenomena} and
  {Bodies}. Space Science Reviews 84, 327--471.
\newline\urlprefix\url{https://ui.adsabs.harvard.edu/abs/1998SSRv...84..327C/abstract}

\bibitem[{Ceplecha and Revelle(2005)}]{ceplecha_fragmentation_2005}
Ceplecha, Z., Revelle, D.~O., 2005. Fragmentation model of meteoroid motion,
  mass loss, and radiation in the atmosphere. Meteoritics and Planetary Science
  40~(1), 35.
\newline\urlprefix\url{http://adsabs.harvard.edu/abs/2005M%26PS...40...35C}

\bibitem[{Clifton(1973)}]{clifton_television_1973}
Clifton, K.~S., 1973. Television studies of faint meteors. Journal of
  Geophysical Research 78, 6511.
\newline\urlprefix\url{https://ui.adsabs.harvard.edu/abs/1973JGR....78.6511C/abstract}

\bibitem[{Close et~al.(2007)Close, Brown, Campbell-Brown, Oppenheim, and
  Colestock}]{close_meteor_2007}
Close, S., Brown, P., Campbell-Brown, M., Oppenheim, M., Colestock, P., 2007.
  Meteor head echo radar data: {Mass}-velocity selection effects. Icarus 186,
  547--556.
\newline\urlprefix\url{https://ui.adsabs.harvard.edu/abs/2007Icar..186..547C/abstract}

\bibitem[{Close et~al.(2000)Close, Hunt, Minardi, and
  McKeen}]{close_analysis_2000}
Close, S., Hunt, S.~M., Minardi, M.~J., McKeen, F.~M., 2000. Analysis of
  {Perseid} meteor head echo data collected using the {Advanced} {Research}
  {Projects} {Agency} {Long}-{Range} {Tracking} and {Instrumentation} {Radar}
  ({ALTAIR}). Radio Science 35, 1233.
\newline\urlprefix\url{https://ui.adsabs.harvard.edu/abs/2000RaSc...35.1233C/abstract}

\bibitem[{Close et~al.(2005)Close, Oppenheim, Durand, and
  Dyrud}]{close_new_2005}
Close, S., Oppenheim, M., Durand, D., Dyrud, L., 2005. A new method for
  determining meteoroid mass from head echo data. Journal of Geophysical
  Research (Space Physics) 110, A09308.
\newline\urlprefix\url{http://adsabs.harvard.edu/abs/2005JGRA..110.9308C}

\bibitem[{Close et~al.(2004)Close, Oppenheim, Hunt, and
  Coster}]{close_technique_2004}
Close, S., Oppenheim, M., Hunt, S., Coster, A., 2004. A technique for
  calculating meteor plasma density and meteoroid mass from radar head echo
  scattering. Icarus 168, 43--52.
\newline\urlprefix\url{http://adsabs.harvard.edu/abs/2004Icar..168...43C}

\bibitem[{Cook et~al.(1980)Cook, Weekes, Williams, and
  Omongain}]{cook_flux_1980}
Cook, A.~F., Weekes, T.~C., Williams, J.~T., Omongain, E., 1980. Flux of
  optical meteors down to {MPG} = +12. Monthly Notices of the Royal
  Astronomical Society 193, 645--666.
\newline\urlprefix\url{https://ui.adsabs.harvard.edu/abs/1980MNRAS.193..645C/abstract}

\bibitem[{Fentzke et~al.(2009)Fentzke, Janches, and
  Sparks}]{fentzke_latitudinal_2009}
Fentzke, J.~T., Janches, D., Sparks, J.~J., 2009. Latitudinal and seasonal
  variability of the micrometeor input function: {A} study using model
  predictions and observations from {Arecibo} and {PFISR}. Journal of
  Atmospheric and Solar-Terrestrial Physics 71, 653--661.
\newline\urlprefix\url{https://ui.adsabs.harvard.edu/abs/2009JASTP..71..653F/abstract}

\bibitem[{Flynn(2002)}]{flynn_extraterrestrial_2002}
Flynn, G.~J., 2002. Extraterrestrial {Dust} in the {Near}-{Earth}
  {Environment}. In: Meteors in the {Earth}'s {Atmosphere}. p.~77.
\newline\urlprefix\url{https://ui.adsabs.harvard.edu/abs/2002mea..book...77F/abstract}

\bibitem[{Friichtenicht and Becker(1973)}]{friichtenicht_determination_1973}
Friichtenicht, J.~F., Becker, D.~G., 1973. Determination of {Meteor}
  {Parameters} {Using} {Laboratory} {Simulation} {Techniques}. NASA Special
  Publication 319, 53.
\newline\urlprefix\url{http://adsabs.harvard.edu/abs/1973NASSP.319...53F}

\bibitem[{Fujiwara et~al.(1995)Fujiwara, Ueda, Nakamura, and
  Tsutsumi}]{fujiwara_simultaneous_1995}
Fujiwara, Y., Ueda, M., Nakamura, T., Tsutsumi, M., 1995. Simultaneous
  {Observations} of {Meteors} with the {Radar} and {TV} {Systems}. Earth Moon
  and Planets 68, 277--282.
\newline\urlprefix\url{https://ui.adsabs.harvard.edu/abs/1995EM%26P...68..277F/abstract}

\bibitem[{Gruen et~al.(1992)Gruen, Fechtig, Kissel, Linkert, Maas, McDonnell,
  Morfill, Schwehm, Zook, and Giese}]{gruen_ulysses_1992}
Gruen, E., Fechtig, H., Kissel, J., Linkert, D., Maas, D., McDonnell, J. A.~M.,
  Morfill, G.~E., Schwehm, G., Zook, H.~A., Giese, R.~H., 1992. The {ULYSSES}
  dust experiment. Astronomy and Astrophysics Supplement Series 92, 411--423.
\newline\urlprefix\url{https://ui.adsabs.harvard.edu/abs/1992A%26AS...92..411G/abstract}

\bibitem[{Gr{\"u}n et~al.(1985)Gr{\"u}n, Zook, Fechtig, and
  Giese}]{grun_collisional_1985}
Gr{\"u}n, E., Zook, H.~A., Fechtig, H., Giese, R.~H., 1985. Collisional balance
  of the meteoritic complex. Icarus 62, 244--272.
\newline\urlprefix\url{https://ui.adsabs.harvard.edu/abs/1985Icar...62..244G/abstract}

\bibitem[{Gural et~al.(2014)Gural, {\v{S}}egon, Andrei{\'c}, Skoki{\'c},
  Korlevi{\'c}, Vida, Novoselnik, and Gostinski}]{gural_results_2014}
Gural, P., {\v{S}}egon, D., Andrei{\'c}, {\v{Z}}., Skoki{\'c}, I.,
  Korlevi{\'c}, K., Vida, D., Novoselnik, F., Gostinski, D., 2014. Results of
  {CMN} 2013 search for new showers across {CMN} and {SonotaCo} databases {II}.
  WGN, Journal of the International Meteor Organization 42, 132--138.
\newline\urlprefix\url{http://adsabs.harvard.edu/abs/2014JIMO...42..132G}

\bibitem[{Gurnett et~al.(1997)Gurnett, Ansher, Kurth, and
  Granroth}]{gurnett_micron-sized_1997}
Gurnett, D.~A., Ansher, J.~A., Kurth, W.~S., Granroth, L.~J., 1997.
  Micron-sized dust particles detected in the outer solar system by the
  {Voyager} 1 and 2 plasma wave instruments. Geophysical Research Letters 24,
  3125--3128.
\newline\urlprefix\url{https://ui.adsabs.harvard.edu/abs/1997GeoRL..24.3125G/abstract}

\bibitem[{Hawkes and Jones(1975{\natexlab{a}})}]{hawkes_quantitative_1975}
Hawkes, R.~L., Jones, J., 1975{\natexlab{a}}. A quantitative model for the
  ablation of dustball meteors. Monthly Notices of the Royal Astronomical
  Society 173, 339--356.
\newline\urlprefix\url{https://ui.adsabs.harvard.edu/abs/1975MNRAS.173..339H/abstract}

\bibitem[{Hawkes and Jones(1975{\natexlab{b}})}]{hawkes_television_1975}
Hawkes, R.~L., Jones, J., 1975{\natexlab{b}}. Television observations of faint
  meteors. {I} - {Mass} distribution and diurnal rate variation. Monthly
  Notices of the Royal Astronomical Society 170, 363--377.
\newline\urlprefix\url{https://ui.adsabs.harvard.edu/abs/1975MNRAS.170..363H/abstract}

\bibitem[{Hawkins(1956{\natexlab{a}})}]{hawkins_radio_1956}
Hawkins, G.~S., 1956{\natexlab{a}}. A radio echo survey of sporadic meteor
  radiants. Monthly Notices of the Royal Astronomical Society 116, 92.
\newline\urlprefix\url{https://ui.adsabs.harvard.edu/1956MNRAS.116...92H/abstract}

\bibitem[{Hawkins(1956{\natexlab{b}})}]{hawkins_variation_1956}
Hawkins, G.~S., 1956{\natexlab{b}}. Variation in the occurrence rate of
  meteors. The Astronomical Journal 61, 386.
\newline\urlprefix\url{https://ui.adsabs.harvard.edu/abs/1956AJ.....61..386H/abstract}

\bibitem[{Hawkins and Upton(1958)}]{hawkins_influx_1958}
Hawkins, G.~S., Upton, E. K.~L., 1958. The {Influx} {Rate} of {Meteors} in the
  {Earth}'s {Atmosphere}. The Astrophysical Journal 128, 727.
\newline\urlprefix\url{https://ui.adsabs.harvard.edu/abs/1958ApJ...128..727H/abstract}

\bibitem[{Hill et~al.(2005)Hill, Rogers, and Hawkes}]{hill_high_2005}
Hill, K.~A., Rogers, L.~A., Hawkes, R.~L., 2005. High geocentric velocity
  meteor ablation. Astronomy and Astrophysics 444, 615--624.
\newline\urlprefix\url{http://adsabs.harvard.edu/abs/2005A%26A...444..615H}

\bibitem[{Hughes(1974)}]{hughes_influx_1974}
Hughes, D.~W., 1974. The influx of visual sporadic meteors. Monthly Notices of
  the Royal Astronomical Society 166, 339.
\newline\urlprefix\url{https://ui.adsabs.harvard.edu/1974MNRAS.166..339H/abstract}

\bibitem[{Hughes and Stephenson(1972)}]{hughes_diurnal_1972}
Hughes, D.~W., Stephenson, D.~G., 1972. The diurnal variation in the
  massdistribution of sporadic meteors. Monthly Notices of the Royal
  Astronomical Society 155, 403.
\newline\urlprefix\url{https://ui.adsabs.harvard.edu/1972MNRAS.155..403H/abstract}

\bibitem[{Iye et~al.(2007)Iye, Tanaka, Yanagisawa, Ebizuka, Ohnishi, Hirose,
  Asami, Komiyama, and Furusawa}]{iye_suprimecam_2007}
Iye, M., Tanaka, M., Yanagisawa, M., Ebizuka, N., Ohnishi, K., Hirose, C.,
  Asami, N., Komiyama, Y., Furusawa, H., 2007. {SuprimeCam} {Observation} of
  {Sporadic} {Meteors} during {Perseids} 2004. Publications of the Astronomical
  Society of Japan 59, 841--855.
\newline\urlprefix\url{https://ui.adsabs.harvard.edu/abs/2007PASJ...59..841I/abstract}

\bibitem[{Janches et~al.(2019)Janches, Brunini, and
  Hormaechea}]{janches_decade_2019}
Janches, D., Brunini, C., Hormaechea, J.~L., 2019. A {Decade} of {Sporadic}
  {Meteoroid} {Mass} {Distribution} {Indices} in the {Southern} {Hemisphere}
  {Derived} from {SAAMER}’s {Meteor} {Observations}. The Astronomical Journal
  157~(6), 240.
\newline\urlprefix\url{http://adsabs.harvard.edu/abs/2019AJ....157..240J}

\bibitem[{Janches et~al.(2015)Janches, Close, Hormaechea, Swarnalingam, Murphy,
  O'Connor, Vandepeer, Fuller, Fritts, and Brunini}]{janches_southern_2015}
Janches, D., Close, S., Hormaechea, J.~L., Swarnalingam, N., Murphy, A.,
  O'Connor, D., Vandepeer, B., Fuller, B., Fritts, D.~C., Brunini, C., 2015.
  The {Southern} {Argentina} {Agile} {MEteor} {Radar} {Orbital} {System}
  ({SAAMER}-{OS}): {An} {Initial} {Sporadic} {Meteoroid} {Orbital} {Survey} in
  the {Southern} {Sky}. The Astrophysical Journal 809, 36.
\newline\urlprefix\url{https://ui.adsabs.harvard.edu/abs/2015ApJ...809...36J/abstract}

\bibitem[{Janches et~al.(2014)Janches, Hocking, Pifko, Hormaechea, Fritts,
  Brunini, Michell, and Samara}]{janches_interferometric_2014}
Janches, D., Hocking, W., Pifko, S., Hormaechea, J.~L., Fritts, D.~C., Brunini,
  C., Michell, R., Samara, M., 2014. Interferometric meteor head echo
  observations using the {Southern} {Argentina} {Agile} {Meteor} {Radar}.
  Journal of Geophysical Research (Space Physics) 119, 2269--2287.
\newline\urlprefix\url{http://adsabs.harvard.edu/abs/2014JGRA..119.2269J}

\bibitem[{Jenniskens(2017)}]{jenniskens_meteor_2017}
Jenniskens, P., 2017. Meteor showers in review. Planetary and Space Science
  143, 116--124.
\newline\urlprefix\url{http://www.sciencedirect.com/science/article/pii/S0032063316303579}

\bibitem[{Jenniskens et~al.(2011)Jenniskens, Gural, Dynneson, Grigsby, Newman,
  Borden, Koop, and Holman}]{jenniskens_cams:_2011}
Jenniskens, P., Gural, P.~S., Dynneson, L., Grigsby, B.~J., Newman, K.~E.,
  Borden, M., Koop, M., Holman, D., 2011. {CAMS}: {Cameras} for {Allsky}
  {Meteor} {Surveillance} to establish minor meteor showers. Icarus 216, 40.
\newline\urlprefix\url{https://ui.adsabs.harvard.edu/2011Icar..216...40J/abstract}

\bibitem[{Jenniskens and N{\'e}non(2016)}]{jenniskens_cams_2016}
Jenniskens, P., N{\'e}non, Q., 2016. {CAMS} verification of single-linked
  high-threshold {D}-criterion detected meteor showers. Icarus 266, 371--383.
\newline\urlprefix\url{https://ui.adsabs.harvard.edu/abs/2016Icar..266..371J/abstract}

\bibitem[{Jones(1997)}]{jones_theoretical_1997}
Jones, W., 1997. Theoretical and observational determinations of the ionization
  coefficient of meteors. Monthly Notices of the Royal Astronomical Society
  288, 995.
\newline\urlprefix\url{https://ui.adsabs.harvard.edu/1997MNRAS.288..995J/abstract}

\bibitem[{Kanamori(2009)}]{kanamori_meteor_2009}
Kanamori, T., 2009. A meteor shower catalog based on video observations in
  2007-2008. WGN, Journal of the International Meteor Organization 37, 55.
\newline\urlprefix\url{https://ui.adsabs.harvard.edu/2009JIMO...37...55S/abstract}

\bibitem[{Kero et~al.(2011)Kero, Szasz, Nakamura, Meisel, Ueda, Fujiwara,
  Terasawa, Miyamoto, and Nishimura}]{kero_first_2011}
Kero, J., Szasz, C., Nakamura, T., Meisel, D.~D., Ueda, M., Fujiwara, Y.,
  Terasawa, T., Miyamoto, H., Nishimura, K., 2011. First results from the
  2009-2010 {MU} radar head echo observation programme for sporadic and shower
  meteors: the {Orionids} 2009. Monthly Notices of the Royal Astronomical
  Society 416, 2550--2559.
\newline\urlprefix\url{https://ui.adsabs.harvard.edu/abs/2011MNRAS.416.2550K/abstract}

\bibitem[{Kero et~al.(2012{\natexlab{a}})Kero, Szasz, Nakamura, Meisel, Ueda,
  Fujiwara, Terasawa, Nishimura, and Watanabe}]{kero_2009-2010_2012}
Kero, J., Szasz, C., Nakamura, T., Meisel, D.~D., Ueda, M., Fujiwara, Y.,
  Terasawa, T., Nishimura, K., Watanabe, J., 2012{\natexlab{a}}. The 2009-2010
  {MU} radar head echo observation programme for sporadic and shower meteors:
  radiant densities and diurnal rates. Monthly Notices of the Royal
  Astronomical Society 425~(1), 135--146.
\newline\urlprefix\url{https://ui.adsabs.harvard.edu/abs/2012MNRAS.425..135K/abstract}

\bibitem[{Kero et~al.(2012{\natexlab{b}})Kero, Szasz, Nakamura, Terasawa,
  Miyamoto, and Nishimura}]{kero_meteor_2012}
Kero, J., Szasz, C., Nakamura, T., Terasawa, T., Miyamoto, H., Nishimura, K.,
  2012{\natexlab{b}}. A meteor head echo analysis algorithm for the lower {VHF}
  band. Annales Geophysicae 30~(4), 639.
\newline\urlprefix\url{https://ui.adsabs.harvard.edu/abs/2012AnGeo..30..639K/abstract}

\bibitem[{Kero et~al.(2008)Kero, Szasz, Pellinen-Wannberg, Wannberg, Westman,
  and Meisel}]{kero_three-dimensional_2008}
Kero, J., Szasz, C., Pellinen-Wannberg, A., Wannberg, G., Westman, A., Meisel,
  D.~D., 2008. Three-dimensional radar observation of a submillimeter meteoroid
  fragmentation. Geophysical Research Letters 35, L04101.
\newline\urlprefix\url{http://adsabs.harvard.edu/abs/2008GeoRL..35.4101K}

\bibitem[{Kojima et~al.(2018)Kojima, Sako, Ohsawa, Takahashi, Doi, Kobayashi,
  Aoki, Arima, Arimatsu, Ichiki, Ikeda, Inooka, Ita, Kasuga, Kokubo, Konishi,
  Maehara, Matsunaga, Mitsuda, Miyata, Mori, Morii, Morokuma, Motohara, Nakada,
  Okumura, Sarugaku, Sato, Shigeyama, Soyano, Tanaka, Tarusawa, Tominaga,
  Totani, Urakawa, Usui, Watanabe, Yamashita, and
  Yoshikawa}]{kojima_evaluation_2018}
Kojima, Y., Sako, S., Ohsawa, R., Takahashi, H., Doi, M., Kobayashi, N., Aoki,
  T., Arima, N., Arimatsu, K., Ichiki, M., Ikeda, S., Inooka, K., Ita, Y.,
  Kasuga, T., Kokubo, M., Konishi, M., Maehara, H., Matsunaga, N., Mitsuda, K.,
  Miyata, T., Mori, Y., Morii, M., Morokuma, T., Motohara, K., Nakada, Y.,
  Okumura, S.-I., Sarugaku, Y., Sato, M., Shigeyama, T., Soyano, T., Tanaka,
  M., Tarusawa, K., Tominaga, N., Totani, T., Urakawa, S., Usui, F., Watanabe,
  J., Yamashita, T., Yoshikawa, M., 2018. Evaluation of large pixel {CMOS}
  image sensors for the {Tomo}-e {Gozen} wide field camera. In: Proc. {SPIE}.
  Vol. 10709. International Society for Optics and Photonics, p. 107091T.
\newline\urlprefix\url{https://doi.org/10.1117/12.2311301}

\bibitem[{Kornos et~al.(2013)Kornos, Koukal, Piffl, and
  Toth}]{kornos_database_2013}
Kornos, L., Koukal, J., Piffl, R., Toth, J., 2013. Database of meteoroid orbits
  from several {European} video networks. In: Proceedings of the 31st
  {International} {Meteor} {Conference}. The International Meteor Organization,
  pp. 21--25.
\newline\urlprefix\url{https://ui.adsabs.harvard.edu/abs/2013pimo.conf...21K/abstract}

\bibitem[{Kres{\'a}kov{\'a}(1966)}]{kresakova_magnitude_1966}
Kres{\'a}kov{\'a}, M., 1966. The {Magnitude} {Distribution} of {Meteors} in
  {Meteor} {Streams}. Contributions of the Astronomical Observatory Skalnate
  Pleso 3, 75.
\newline\urlprefix\url{https://ui.adsabs.harvard.edu/abs/1966CoSka...3...75K/abstract}

\bibitem[{Love and Brownlee(1993)}]{love_direct_1993}
Love, S.~G., Brownlee, D.~E., 1993. A {Direct} {Measurement} of the
  {Terrestrial} {Mass} {Accretion} {Rate} of {Cosmic} {Dust}. Science 262,
  550--553.
\newline\urlprefix\url{http://adsabs.harvard.edu/abs/1993Sci...262..550L}

\bibitem[{Michell(2010)}]{michell_simultaneous_2010}
Michell, R.~G., 2010. Simultaneous optical and radar measurements of meteors
  using the {Poker} {Flat} {Incoherent} {Scatter} {Radar}. Journal of
  Atmospheric and Solar-Terrestrial Physics 72~(16), 1212--1220.
\newline\urlprefix\url{https://ui.adsabs.harvard.edu/abs/2010JASTP..72.1212M/abstract}

\bibitem[{Michell et~al.(2019)Michell, DeLuca, Janches, Chen, and
  Samara}]{michell_simultaneous_2019}
Michell, R.~G., DeLuca, M., Janches, D., Chen, R., Samara, M., 2019.
  Simultaneous optical and dual-frequency radar observations of small mass
  meteors at {Arecibo}. Planetary and Space Science 166, 1--8.
\newline\urlprefix\url{http://adsabs.harvard.edu/abs/2019P%26SS..166....1M}

\bibitem[{Michell et~al.(2015)Michell, Janches, Samara, Hormaechea, Brunini,
  and Bibbo}]{michell_simultaneous_2015}
Michell, R.~G., Janches, D., Samara, M., Hormaechea, J.~L., Brunini, C., Bibbo,
  I., 2015. Simultaneous optical and radar observations of meteor head-echoes
  utilizing {SAAMER}. Planetary and Space Science 118, 95--101.
\newline\urlprefix\url{https://ui.adsabs.harvard.edu/abs/2015P%26SS..118...95M/abstract}

\bibitem[{Myers et~al.(2001)Myers, Sande, Miller, Warren, and
  Tracewell}]{myers_vizier_2001}
Myers, J.~R., Sande, C.~B., Miller, A.~C., Warren, W.~H., Tracewell, D.~A.,
  2001. {VizieR} {Online} {Data} {Catalog}: {SKY2000} {Catalog}, {Version} 4
  ({Myers}+ 2002). VizieR Online Data Catalog, V/109.
\newline\urlprefix\url{https://ui.adsabs.harvard.edu/abs/2001yCat.5109....0M/abstract}

\bibitem[{Nishimura et~al.(2001)Nishimura, Sato, Nakamura, and
  Ueda}]{nishimura_high_2001}
Nishimura, K., Sato, T., Nakamura, T., Ueda, M., 2001. High sensitivity
  radar-optical observations of faint meteors. IEICE Transactions on
  Electronics E84-C~(12), 1877--1884.

\bibitem[{Nishimura et~al.(2002)Nishimura, Ohnishi, Dobashi, Watanabe, Miyata,
  and Nakada}]{nishimura_optical_2002}
Nishimura, S., Ohnishi, K., Dobashi, K., Watanabe, J.-I., Miyata, T., Nakada,
  Y., 2002. Optical {Imaging} of the {Radiant} {Points} of {Leonids} during the
  2001 {Storm} with the 105cm {Kiso} {Schmidt} {Telescope}. Publications of the
  Astronomical Society of Japan 54, L83--L88.
\newline\urlprefix\url{https://ui.adsabs.harvard.edu/abs/2002PASJ...54L..83N/abstract}

\bibitem[{Ohsawa et~al.(2019)Ohsawa, Sako, Sarugaku, Usui, Ootsubo, Fujiwara,
  Sato, Kasuga, Arimatsu, Watanabe, Doi, Kobayashi, Takahashi, Motohara,
  Morokuma, Konishi, Aoki, Soyano, Tarusawa, Mori, Nakada, Ichiki, Arima,
  Kojima, Morita, Shigeyama, Ita, Kokubo, Mitsuda, Maehara, Tominaga,
  Yamashita, Ikeda, Morii, Urakawa, Okumura, and
  Yoshikawa}]{ohsawa_luminosity_2019}
Ohsawa, R., Sako, S., Sarugaku, Y., Usui, F., Ootsubo, T., Fujiwara, Y., Sato,
  M., Kasuga, T., Arimatsu, K., Watanabe, J.-i., Doi, M., Kobayashi, N.,
  Takahashi, H., Motohara, K., Morokuma, T., Konishi, M., Aoki, T., Soyano, T.,
  Tarusawa, K., Mori, Y., Nakada, Y., Ichiki, M., Arima, N., Kojima, Y.,
  Morita, M., Shigeyama, T., Ita, Y., Kokubo, M., Mitsuda, K., Maehara, H.,
  Tominaga, N., Yamashita, T., Ikeda, S., Morii, M., Urakawa, S., Okumura,
  S.-i., Yoshikawa, M., 2019. Luminosity function of faint sporadic meteors
  measured with a wide-field {CMOS} mosaic camera {Tomo}-e {PM}. Planetary and
  Space Science 165, 281--292.
\newline\urlprefix\url{https://ui.adsabs.harvard.edu/abs/2019P%26SS..165..281O/abstract}

\bibitem[{Ohsawa et~al.(2016)Ohsawa, Sako, Takahashi, Kikuchi, Doi, Kobayashi,
  Aoki, Arimatsu, Ichiki, Ikeda, Ita, Kasuga, Kawakita, Kokubo, Maehara,
  Matsunaga, Mito, Mitsuda, Miyata, Mori, Mori, Morii, Morokuma, Motohara,
  Nakada, Okumura, Onozato, Osawa, Sarugaku, Sato, Shigeyama, Soyano, Tanaka,
  Taniguchi, Tanikawa, Tarusawa, Tominaga, Totani, Urakawa, Usui, Watanabe,
  Yamaguchi, and Yoshikawa}]{ohsawa_development_2016}
Ohsawa, R., Sako, S., Takahashi, H., Kikuchi, Y., Doi, M., Kobayashi, N., Aoki,
  T., Arimatsu, K., Ichiki, M., Ikeda, S., Ita, Y., Kasuga, T., Kawakita, H.,
  Kokubo, M., Maehara, H., Matsunaga, N., Mito, H., Mitsuda, K., Miyata, T.,
  Mori, K., Mori, Y., Morii, M., Morokuma, T., Motohara, K., Nakada, Y.,
  Okumura, S.-i., Onozato, H., Osawa, K., Sarugaku, Y., Sato, M., Shigeyama,
  T., Soyano, T., Tanaka, M., Taniguchi, Y., Tanikawa, A., Tarusawa, K.,
  Tominaga, N., Totani, T., Urakawa, S., Usui, F., Watanabe, J., Yamaguchi, J.,
  Yoshikawa, M., 2016. Development of a real-time data processing system for a
  prototype of the {Tomo}-e {Gozen} wide field {CMOS} camera. In: Proc. {SPIE}.
  Vol. 9913. International Society for Optics and Photonics, p. 991339.
\newline\urlprefix\url{http://proceedings.spiedigitallibrary.org/proceeding.aspx?articleid=2543636}

\bibitem[{Pellinen-Wannberg and Wannberg(1994)}]{pellinen-wannberg_meteor_1994}
Pellinen-Wannberg, A., Wannberg, G., 1994. Meteor observations with the
  {European} incoherent scatter {UHF} radar. Journal of Geophysical Research
  99, 11379.
\newline\urlprefix\url{https://ui.adsabs.harvard.edu/1994JGR....9911379P/abstract}

\bibitem[{Pellinen-Wannberg et~al.(1998)Pellinen-Wannberg, Westman, Wannberg,
  and Kaila}]{pellinen-wannberg_meteor_1998}
Pellinen-Wannberg, A., Westman, A., Wannberg, G., Kaila, K., 1998. Meteor
  fluxes and visual magnitudes from {EISCAT} radar event rates: a comparison
  with cross-section based magnitude estimates and optical data. Annales
  Geophysicae 16~(11), 1475--1485.
\newline\urlprefix\url{https://www.ann-geophys.net/16/1475/1998/}

\bibitem[{Plane(2012)}]{plane_cosmic_2012}
Plane, J. M.~C., 2012. Cosmic dust in the earth's atmosphere. Chemical Society
  Reviews, Vol. 41, p. 6507-6518, 2012 41, 6507--6518.
\newline\urlprefix\url{https://ui.adsabs.harvard.edu/abs/2012ChSRv..41.6507P/abstract}

\bibitem[{Popova(2004)}]{popova_meteoroid_2004}
Popova, O., 2004. Meteoroid ablation models. Earth Moon and Planets 95,
  303--319.
\newline\urlprefix\url{https://ui.adsabs.harvard.edu/abs/2004EM%26P...95..303P/abstract}

\bibitem[{Rendtel(2004)}]{rendtel_population_2004}
Rendtel, J., 2004. The population index of sporadic meteors. In: Proceedings of
  the 22nd {International} {Meteor} {Conference}. Vol.~22. Bollmannsruh,
  Germany, pp. 114--122.
\newline\urlprefix\url{https://ui.adsabs.harvard.edu/abs/2004pimo.conf..114R/abstract}

\bibitem[{Saidov and Simek(1989)}]{saidov_luminous_1989}
Saidov, K.~H., Simek, M., 1989. Luminous {Efficiency} {Coefficient} from
  {Simultaneous} {Meteor} {Observations}. Bulletin of the Astronomical
  Institutes of Czechoslovakia 40, 330.
\newline\urlprefix\url{https://ui.adsabs.harvard.edu/1989BAICz..40..330S/abstract}

\bibitem[{Sako et~al.(2016)Sako, Ohsawa, Takahashi, Kikuchi, Doi, Kobayashi,
  Aoki, Arimatsu, Ichiki, Ikeda, Ita, Kasuga, Kawakita, Kokubo, Maehara,
  Matsunaga, Mito, Mitsuda, Miyata, Mori, Mori, Morii, Morokuma, Motohara,
  Nakada, Osawa, Okumura, Onozato, Sarugaku, Sato, Shigeyama, Soyano, Tanaka,
  Taniguchi, Tanikawa, Tarusawa, Tominaga, Totani, Urakawa, Usui, Watanabe,
  Yamaguchi, and Yoshikawa}]{sako_development_2016}
Sako, S., Ohsawa, R., Takahashi, H., Kikuchi, Y., Doi, M., Kobayashi, N., Aoki,
  T., Arimatsu, K., Ichiki, M., Ikeda, S., Ita, Y., Kasuga, T., Kawakita, H.,
  Kokubo, M., Maehara, H., Matsunaga, N., Mito, H., Mitsuda, K., Miyata, T.,
  Mori, K., Mori, Y., Morii, M., Morokuma, T., Motohara, K., Nakada, Y., Osawa,
  K., Okumura, S.-i., Onozato, H., Sarugaku, Y., Sato, M., Shigeyama, T.,
  Soyano, T., Tanaka, M., Taniguchi, Y., Tanikawa, A., Tarusawa, K., Tominaga,
  N., Totani, T., Urakawa, S., Usui, F., Watanabe, J., Yamaguchi, J.,
  Yoshikawa, M., 2016. Development of a prototype of the {Tomo}-e {Gozen}
  wide-field {CMOS} camera. In: Proc. {SPIE}. Vol. 9908. International Society
  for Optics and Photonics, p. 99083P.
\newline\urlprefix\url{http://proceedings.spiedigitallibrary.org/proceeding.aspx?articleid=2544194}

\bibitem[{Sako et~al.(2018)Sako, Ohsawa, Takahashi, Kojima, Doi, Kobayashi,
  Aoki, Arima, Arimatsu, Ichiki, Ikeda, Inooka, Ita, Kasuga, Kokubo, Konishi,
  Maehara, Matsunaga, Mitsuda, Miyata, Mori, Morii, Morokuma, Motohara, Nakada,
  Okumura, Sarugaku, Sato, Shigeyama, Soyano, Tanaka, Tarusawa, Tominaga,
  Totani, Urakawa, Usui, Watanabe, Yamashita, and Yoshikawa}]{sako_tomo-e_2018}
Sako, S., Ohsawa, R., Takahashi, H., Kojima, Y., Doi, M., Kobayashi, N., Aoki,
  T., Arima, N., Arimatsu, K., Ichiki, M., Ikeda, S., Inooka, K., Ita, Y.,
  Kasuga, T., Kokubo, M., Konishi, M., Maehara, H., Matsunaga, N., Mitsuda, K.,
  Miyata, T., Mori, Y., Morii, M., Morokuma, T., Motohara, K., Nakada, Y.,
  Okumura, S.-I., Sarugaku, Y., Sato, M., Shigeyama, T., Soyano, T., Tanaka,
  M., Tarusawa, K., Tominaga, N., Totani, T., Urakawa, S., Usui, F., Watanabe,
  J., Yamashita, T., Yoshikawa, M., 2018. The {Tomo}-e {Gozen} wide field
  {CMOS} camera for the {Kiso} {Schmidt} telescope. In: Proc. {SPIE}. Vol.
  10702. International Society for Optics and Photonics, p. 107020J.
\newline\urlprefix\url{https://doi.org/10.1117/12.2310049}

\bibitem[{Sato et~al.(2000)Sato, Nakamura, and Nishimura}]{sato_orbit_2000}
Sato, T., Nakamura, T., Nishimura, K., 2000. Orbit {Determination} of {Meteors}
  {Using} the {MU} {Radar}. IEICE TRANSACTIONS on Communications E83-B~(9),
  1990--1995.
\newline\urlprefix\url{http://search.ieice.org/bin/summary.php?id=e83-b_9_1990&category=B&year=2000&lang=E&abst=}

\bibitem[{Schult et~al.(2017)Schult, Stober, Janches, and
  Chau}]{schult_results_2017}
Schult, C., Stober, G., Janches, D., Chau, J.~L., 2017. Results of the first
  continuous meteor head echo survey at polar latitudes. Icarus 297, 1--13.
\newline\urlprefix\url{http://adsabs.harvard.edu/abs/2017Icar..297....1S}

\bibitem[{{\v{S}}egon et~al.(2014){\v{S}}egon, Andrei{\'c}, Gural, Skoki{\'c},
  Korlevi{\'c}, Vida, Novoselnik, and Gostinski}]{segon_results_2014}
{\v{S}}egon, D., Andrei{\'c}, {\v{Z}}., Gural, P., Skoki{\'c}, I.,
  Korlevi{\'c}, K., Vida, D., Novoselnik, F., Gostinski, D., 2014. Results of
  {CMN} 2013 search for new showers across {CMN} and {SonotaCo} databases
  {III}. WGN, Journal of the International Meteor Organization 42, 227--233.
\newline\urlprefix\url{http://adsabs.harvard.edu/abs/2014JIMO...42..227S}

\bibitem[{Southworth and Hawkins(1963)}]{southworth_statistics_1963}
Southworth, R.~B., Hawkins, G.~S., 1963. Statistics of meteor streams.
  Smithsonian Contributions to Astrophysics 7, 261.
\newline\urlprefix\url{https://ui.adsabs.harvard.edu/1963SCoA....7..261S/abstract}

\bibitem[{Sparks et~al.(2009)Sparks, Janches, Nicolls, and
  Heinselman}]{sparks_seasonal_2009}
Sparks, J.~J., Janches, D., Nicolls, M.~J., Heinselman, C.~J., 2009. Seasonal
  and diurnal variability of the meteor flux at high latitudes observed using
  {PFISR}. Journal of Atmospheric and Solar-Terrestrial Physics 71, 644--652.
\newline\urlprefix\url{http://adsabs.harvard.edu/abs/2009JASTP..71..644S}

\bibitem[{{\v{S}}tohl(1968)}]{stohl_seasonal_1968}
{\v{S}}tohl, J., 1968. Seasonal {Variation} in the {Radiant} {Distribution} of
  {Meteors}. Physics and Dynamics of Meteors 33, 298.
\newline\urlprefix\url{https://ui.adsabs.harvard.edu/abs/1968IAUS...33..298S/abstract}

\bibitem[{{\v{S}}tohl(1976)}]{stohl_magnitude_1976}
{\v{S}}tohl, J., 1976. The magnitude distribution of sporadic meteors and its
  variations. Contributions of the Astronomical Observatory Skalnate Pleso 7,
  7.
\newline\urlprefix\url{https://ui.adsabs.harvard.edu/abs/1976CoSka...7....7S/abstract}

\bibitem[{Szalay et~al.(2013)Szalay, Piquette, and
  Hor{\'a}nyi}]{szalay_student_2013}
Szalay, J.~R., Piquette, M., Hor{\'a}nyi, M., 2013. The {Student} {Dust}
  {Counter}: {Status} report at 23 {AU}. Earth, Planets, and Space 65,
  1145--1149.
\newline\urlprefix\url{https://ui.adsabs.harvard.edu/abs/2013EP%26S...65.1145S/abstract}

\bibitem[{Thomas et~al.(2016)Thomas, Hor{\'a}nyi, Janches, Munsat, Simolka, and
  Sternovsky}]{thomas_measurements_2016}
Thomas, E., Hor{\'a}nyi, M., Janches, D., Munsat, T., Simolka, J., Sternovsky,
  Z., 2016. Measurements of the ionization coefficient of simulated iron
  micrometeoroids. Geophysical Research Letters 43, 3645--3652.
\newline\urlprefix\url{http://adsabs.harvard.edu/abs/2016GeoRL..43.3645T}

\bibitem[{Verniani(1965)}]{verniani_luminous_1965}
Verniani, F., 1965. On the {Luminous} {Efficiency} of {Meteors}. Smithsonian
  Contributions to Astrophysics 8, 141.
\newline\urlprefix\url{http://adsabs.harvard.edu/abs/1965SCoA....8..141V}

\bibitem[{Weryk and Brown(2012)}]{weryk_simultaneous_2012}
Weryk, R.~J., Brown, P.~G., 2012. Simultaneous radar and video meteors --- {I}:
  {Metric} comparisons. Planetary and Space Science 62~(1), 132.
\newline\urlprefix\url{https://ui.adsabs.harvard.edu/abs/2012P%26SS...62..132W/abstract}

\bibitem[{Weryk and Brown(2013)}]{weryk_simultaneous_2013}
Weryk, R.~J., Brown, P.~G., 2013. Simultaneous radar and video meteors ---
  {II}: {Photometry} and ionisation. Planetary and Space Science 81, 32.
\newline\urlprefix\url{https://ui.adsabs.harvard.edu/abs/2013P%26SS...81...32W/abstract}

\bibitem[{Weryk et~al.(2013)Weryk, Campbell-Brown, Wiegert, Brown, Krzeminski,
  and Musci}]{weryk_canadian_2013}
Weryk, R.~J., Campbell-Brown, M.~D., Wiegert, P.~A., Brown, P.~G., Krzeminski,
  Z., Musci, R., 2013. The {Canadian} {Automated} {Meteor} {Observatory}
  ({CAMO}): {System} overview. Icarus 225~(1), 614--622.
\newline\urlprefix\url{http://adsabs.harvard.edu/abs/2013Icar..225..614W}

\bibitem[{Zacharias et~al.(2013)Zacharias, Finch, Girard, Henden, Bartlett,
  Monet, and Zacharias}]{zacharias_fourth_2013}
Zacharias, N., Finch, C.~T., Girard, T.~M., Henden, A., Bartlett, J.~L., Monet,
  D.~G., Zacharias, M.~I., 2013. The {Fourth} {US} {Naval} {Observatory} {CCD}
  {Astrograph} {Catalog} ({UCAC4}). The Astronomical Journal 145, 44.
\newline\urlprefix\url{https://ui.adsabs.harvard.edu/2013AJ....145...44Z/abstract}

\end{thebibliography}
\end{document}